\PassOptionsToPackage{final}{graphicx}
\documentclass[review,nonatbib]{elsarticle}

\makeatletter
\let\c@author\relax
\makeatother

\usepackage[style=numeric-comp,backend=biber,natbib=true,mincitenames=1,maxcitenames=2,bibencoding=utf8,sorting=none,sortcites,maxbibnames=99]{biblatex}
\usepackage{lineno}
\modulolinenumbers[5]

\journal{Science of Computer Programming}

\usepackage[draft=false,pdftex,bookmarks,linktocpage,hidelinks]{hyperref}
\hypersetup{
	pdftitle={Automated Evolution of Feature Logging StatementLevels Using Git Histories and Degree of Interest}
	pdfauthor={Yiming Tang, Allan Spektor, Raffi Khatchadourian, and Mehdi Bagherzadeh},
	pdfsubject={Software Engineering, Programming Languages},
	pdfkeywords={logging, software evolution,software repository mining, source code analysis \& transformation, degree of interest}}

\makeatletter
\let\MYcaption\@makecaption%
\makeatother
\usepackage[font=footnotesize]{subcaption}
\makeatletter
\let\@makecaption\MYcaption%
\makeatother

\usepackage[obeyDraft,colorinlistoftodos]{todonotes}

\newcommand{\addrefRK}{\todo[author=Raffi,color=red!40]{Add reference.}}

\newcommand{\insertrefRKi}[1]{\todo[author=Raffi,inline,color=green!40]{#1}}

\usepackage[group-separator={,},group-minimum-digits={3},detect-all]{siunitx}
\DeclareSIUnit[number-unit-product = ]\percent{\char`\%}

\usepackage{rotating}

\usepackage{nicefrac}
\usepackage{textcomp}
\usepackage{soul}
\usepackage{mathrsfs}
\usepackage{amsmath}
\usepackage{amsfonts}
\usepackage{mathtools}
\usepackage{comment}
\usepackage{ellipsis}
\usepackage{mdframed}

\usepackage{algorithm}
\usepackage{algorithmic}

\graphicspath{{./graphics/}} 

\definecolor{applegreen}{rgb}{0.55, 0.71, 0.0}
\definecolor{yellow-green}{rgb}{0.6, 0.8, 0.2}
\definecolor{emerald}{rgb}{0.31, 0.78, 0.47}
\definecolor{blush}{rgb}{0.87, 0.36, 0.51}
\definecolor{cadmiumred}{rgb}{0.89, 0.0, 0.13}

\usepackage[capitalize]{cleveref}

\crefname{listing}{listing}{listings}
\Crefname{listing}{Listing}{Listings}
\crefname{sublisting}{listing}{listings}
\Crefname{sublisting}{Listing}{Listings}

\crefname{ALC@unique}{line}{lines}
\Crefname{ALC@unique}{Line}{Lines}

\crefname{paragraph}{section}{sections}
\Crefname{paragraph}{Section}{Sections}

\newcommand{\Vlabel}[1]{\label[line]{#1}\hypertarget{#1}{}}
\newcommand{\lref}[1]{\hyperlink{#1}{\FancyVerbLineautorefname~\ref*{#1}}}

\usepackage[shortlabels,inline]{enumitem}

\crefname{section}{\S}{\S}
\Crefname{section}{\S}{\S}
\crefname{paragraph}{\S}{\S}
\Crefname{paragraph}{\S}{\S}
\crefname{figure}{fig.}{fig.}
\Crefname{figure}{Fig.}{Fig.}
\crefname{table}{tab.}{tab.}
\Crefname{table}{Tab.~}{Tab.~}
\crefname{definition}{def.}{def.}
\Crefname{definition}{Def.~}{Def.~}
\crefname{listing}{lst.}{lst.}
\Crefname{listing}{Lst.}{Lst.}
\crefname{sublisting}{lst.}{lst.}
\Crefname{sublisting}{Lst.}{Lst.}

\usepackage[frozencache, draft=false]{minted}
\newminted{diff}{
    style=trac,
}
\newminted{java}{}
\newmintinline{java}{}
\newmintinline{text}{
    fontsize=auto,
}
\setmintedinline{
    style=bw,
    fontsize=\small}
\setminted{%
    style=bw,
    fontsize=\scriptsize,
    autogobble,
    breaklines,
    resetmargins,
    breakindentnchars=2,
    linenos,
    tabsize=2,
    breakafter=.,
    breaksymbol=,
    breakautoindent,
    breakaftersymbolpre=,
    numbersep=3pt,
    escapeinside=||}

\newcommand{\ji}[1]{\javainline{#1}}

\newcommand{\jit}[1]{\mintinline[fontsize=auto]{java}{#1}}

\usepackage{booktabs}
\usepackage{tabulary}
\usepackage{threeparttable}
\usepackage{multirow}

\newcommand{\NumProjects}{\num{18}}

\newcommand{\MLOC}{$\sim$\num{3}}
\newcommand{\logs}{$\sim$\num{4}K}
\newcommand{\logsThousands}{$\sim$\num{4}K}
\newcommand{\analyzedPercentage}{\SI{99.22}{\percent}}
\newcommand{\distributionBefore}{\num{1.75}}
\newcommand{\distributionAfter}{\num{2.05}}
\newcommand{\distributionPercentageIncrease}{\SI{17.14}{\percent}}
\newcommand{\distributionPercentageIncreaseApprox}{\SI{\sim20}{\percent}}
\newcommand{\trans}{753}
\newcommand{\transPercent}{\SI{19}{\percent}}

\newcommand{\AvgRuntimePerLog}{\num{10.66}}
\newcommand{\AvgRuntimePerKLOCChanged}{\num{0.89}}
\newcommand{\catchPercentage}{\SI{3.60}{\percent}}
\newcommand{\ifPercentage}{\SI{10.62}{\percent}}

\newcommand{\condPercentage}{\SI{6.24}{\percent}}
\newcommand{\withKeysPercentage}{\SI{34.23}{\percent}}
\newcommand{\withoutKeysPercentage}{\SI{16.18}{\percent}}

\newcommand{\logsNotAlteredDueToKeywords}{\num{\sim2}K}

\newcommand{\transDist}{\num{2.13}}
\newcommand{\lowPercentage}{\SI{89.51}{\percent}}
\newcommand{\raisePercentage}{\SI{10.36}{\percent}}
\newcommand{\pullRequestedProjects}{\num{16}}

\newcommand{\pullRequestedProjectsAccepted}{\num{2}}
\newcommand{\pullRequestedProjectsAcceptedWord}{two}

\newcommand{\chosenCommitsFromEachContextType}{\num{\sim20}}
\newcommand{\bugsNumberOfSubjects}{\num{12}}

\newcommand{\bugsIdealVersusActualRateApprox}{\SI{\sim83}{\percent}}
\newcommand{\bugsIdealVersusActualFraction}{$\nicefrac{204}{245}$}

\newcommand{\jdp}{\href{https://github.com/iluwatar/java-design-patterns}{\textinline{java-design-patterns}}}

\newcommand{\blueocean}{\href{https://github.com/jenkinsci/blueocean-plugin}{\textinline{blueocean-plugin}}}
\newcommand{\jenkins}{\href{https://github.com/jenkinsci/jenkins}{\textinline{Jenkins}}}
\newcommand{\selenium}{\href{https://github.com/SeleniumHQ/selenium}{\textinline{selenium}}}
\newcommand{\IRCT}{\href{https://github.com/hms-dbmi/IRCT}{\textinline{IRCT}}}
\newcommand{\CoreNLP}{\href{https://github.com/stanfordnlp/CoreNLP}{\textinline{CoreNLP}}}

\newcommand{\deviation}{\href{https://github.com/DeviationTX/deviation-upload.git}{\textinline{deviation-upload}}}

\newcommand{\junit}{\href{https://github.com/junit-team/junit5}{\textinline{junit5}}}
\newcommand{\guava}{\href{https://github.com/google/guava}{\textinline{guava}}}

\newcommand{\gitcproc}{\href{https://github.com/caseycas/gitcproc}{\textinline{gitcproc}}}

\newcounter{takeaway}
\newmdenv[%
    linecolor=black,
    outerlinewidth=0pt,
    skipabove=1pt,
    skipbelow=1pt,
    settings={\global\refstepcounter{takeaway}},
]{mytakeaway}

\newcommand{\takeaway}[1]{
    \begin{mytakeaway}
	\textbf{\textit{Takeaway~\arabic{takeaway}}}: #1
    \end{mytakeaway}
}

\newlist{questions}{enumerate}{2}
\setlist[questions,1]{label=\textbf{RQ\arabic*.},ref=\textbf{RQ\arabic*}}
\setlist[questions,2]{label=(\arabic*),ref=\thequestionsi\textbf{.\arabic*}} 

\begin{document}

\begin{frontmatter}

\title{Automated Evolution of Feature Logging Statement Levels Using Git Histories and Degree of Interest
}
 
 
\author[gc]{Yiming Tang\corref{cor1}}
\ead{ytang3@gradcenter.cuny.edu}
\ead[url]{http://linkedin.com/in/gracetang1993}
\cortext[cor1]{Corresponding author.}

\author[hunter]{Allan Spektor}
\ead{allan.spektor03@myhunter.cuny.edu}
\ead[url]{https://www.linkedin.com/in/allan-spektor}

\author[hunter]{Raffi Khatchadourian}
\ead{raffi.khatchadourian@hunter.cuny.edu}
\ead[url]{http://cs.hunter.cuny.edu/~Raffi.Khatchadourian99}

\author[oakland]{Mehdi Bagherzadeh}
\ead{mbagherzadeh@oakland.edu}
\ead[url]{http://mbagherz.bitbucket.io}

\address[gc]{Department of Computer Science, City University of New York (CUNY) Graduate Center, 365 5th Ave, New York, NY 10016 USA}

\address[hunter]{Department of Computer Science, City University of New York (CUNY) Hunter College, 695 Park Avenue, Room HN 1008, New York, NY 10065 USA}

\address[oakland]{Department of Computer Science \& Engineering, Oakland University, Rochester, MI 48309 USA}


\begin{abstract}

    Logging---used for
    system events and security breaches to describe more informational yet essential aspects of software features---is pervasive.
    Given the high transactionality of today’s software, logging effectiveness can be reduced by information overload. Log levels help alleviate this problem by correlating a priority to logs that can be later filtered. As software evolves, however, levels of logs documenting surrounding feature implementations may also require modification as features once deemed important may have decreased in urgency and vice-versa. 
    We present an automated approach that assists developers in evolving levels of such (feature) logs. The approach, based on mining Git histories and manipulating a degree of interest (DOI) model\footnote{Degree of interest model (DOI) was proposed by~\citet{Kersten2005} to gauge the degree of developers’ interests in program elements.}, transforms source code to revitalize feature log levels based on the ``interestingness'' of the surrounding code.
    Built upon JGit and Mylyn, the approach is implemented as an Eclipse IDE plug-in and evaluated on \NumProjects\ Java projects with \MLOC\ million lines of code and \logsThousands\ log statements. Our tool successfully analyzes \analyzedPercentage\ of logging statements, increases log level distributions by \distributionPercentageIncreaseApprox, 
    and increases the focus of logs in bug fix contexts \bugsIdealVersusActualRateApprox\ of the time. Moreover,
    pull (patch) requests were integrated into large and popular open-source projects.
    The results indicate that the approach is promising in assisting developers in evolving feature log levels.

\end{abstract}


\end{frontmatter}


\section{Introduction}

Modern software
typically includes
logging, which documents useful information about a system's behavior at run-time and facilitates
system understanding. Logs help diagnose run-time issues and can be used to monitor processes~\cite{Rozinat2005}, transfer knowledge~\cite{Kabinna2018}, and detect errors~\cite{Tan2008,Zeng2019,Syer2013}. Other logs, feature logs, may be more informational yet essential as they describe aspects of features the surrounding code implements. 

However, the high transactionality of today’s software can cause logging to be less effective due to information overload. The sheer number of logs emitted can make it challenging to debug during development; logs pertaining to auxiliary features may be tangled with those features under current development. Also, parsing necessary information from logs to understand system behavior, how features interact, and diagnosing problems can be challenging.\addrefRK%

To help alleviate these problems, logging frameworks and libraries empower developers to write logging statements consisting of several parts dictating how the log should be emitted, if at all. A logging statement is comprised of a particular log \emph{object}, 
each of which is associated with a \emph{run-time} level and other attributes. A logging \emph{method} is invoked on the log object; one parameter is a log priority \emph{level}. Log levels are ordered, and---during execution---the log \emph{message} is emitted iff the log statement level is greater than or equal to the log object run-time level. Messages are typically dynamically constructed with static text and dynamic contexts, such as the contents of one or more variables~\cite{Chen2017}. For example, the following statement outputs system-health information iff the run-time level of \javainline{logger} is $\leq$ \javainline{FINER}~\cite{Oracle2018a}: \javainline{logger.log(Level.FINER, "Health is: " + DiagnosisMessages.systemHealthStatus())}. Controlling the log run-time level affords developers the ability to limit the types of log information emitted either for particular environments (e.g., development, deployment) or other factors.

As software evolves, however, levels of logging statements correlated with surrounding feature implementations may also need to be modified.
Such \emph{feature} logging statements could, for example, serve as algorithm checkpoints, where critical variables are outputted for validation and progress is ensured.
Ideally, levels of \emph{feature} logs would \emph{evolve with} systems as they are developed, with higher log levels (e.g., \ji{INFO}) being assigned to logs corresponding to features with more current stakeholder interest than those with less (e.g., \ji{FINEST}). However, as developers tend not to manually change log levels~\cite{Li2017}, feature log levels may become stale, causing irrelevant logs to accumulate, increased information overload, and tangling of relevant feature logs with those not currently being developed, thereby complicating debugging. Furthermore, \emph{manually} maintaining log levels can be tedious and error- and omission-prone as logging statements are highly scattered~\cite{Zeng2019}.
Moreover, developers may not use the full spectrum of available levels.

Existing approaches~\cite{Li2017,Hassani2018,Chen2017,Kabinna2018,He2018} focus on either new logging statements or messages. Logger hierarchies~\cite{Oracle2018a,ASF2020} may be useful
but still require manual maintenance. To the best of our knowledge, there is currently no automated solution for logging statement level evolution. Therefore, we present an automated approach that assists developers in evolving feature logging statement levels. The approach first mines Git repositories to discover developers' ``interestingness'' of code surrounding feature logging statements by adapting the degree of interest (DOI) model of Mylyn~\cite{Kersten2005}. Mylyn~\cite{EclipseFoundation2020} is a standard Eclipse Integrated Development Environment (IDE)~\cite{EclipseFoundation2020a} plug-in that facilitates software evolution by focusing graphical IDE components so that only artifacts related to the currently active task are revealed~\cite{Kersten2006}. Mylyn manipulates DOI so that artifacts (e.g., files) with more interaction are more prominently displayed in the IDE than those less recently used. Each program element is associated with a float value named DOI value that is used to gauge the developers' degree of interest in it. 
The approach later correlates ``interestingness'' of code surrounding feature logging statements with feature log levels. If our approach detects a mismatch between feature log levels and feature interests, it could suggest appropriate log levels to the feature logging statements.

We programmatically manipulate DOI using modifications made in source code repositories. Our approach transforms code to reinvigorate feature logging statement levels, pulling those related to features whose implementations are worked on more and more recently to the forefront, while pushing those worked on less and less recently to the background. Our goal is information overload reduction and improved debugging by automatically bringing more relevant features to developers' attention and vice-versa throughout system evolution.

Logging levels are often used to differentiate various logging \emph{categories}, i.e., levels having special semantics that are not on a ``sliding scale.'' Altering such levels may violate the preservation of the log's intended semantics. In this work, we focus on the levels \emph{feature} logs, i.e., those highly related to feature implementations, as feature interests vary over time and whose related logging statements may benefit from aligning levels correspondingly. Thus, to distinguish feature logs from those that are more categorical, e.g., those conveying more critical information (e.g., errors, security),
a series of novel heuristics, mainly derived from first-hand developer interactions, are introduced. On the other hand, the heuristics also account for less-critical debugging logs, e.g., tracing, using a keyword-based technique. This effort focuses our approach on only manipulating logging statements tied to features to better coordinate them with developers' current interests.

Our approach is implemented as an open-source plug-in to the Eclipse IDE, though it may be used with other IDEs via popular build systems. It supports two popular logging frameworks
and integrates
with JGit~\cite{EclipseFoundation2020b}
and Mylyn.
The evaluation involved
\NumProjects\ Java projects of varying sizes and domains with a total of \MLOC\ million lines of code and \logsThousands\ logging statements. Our study indicates that
\begin{enumerate*}[(i)] 

	\item given its ability to process a significant number and size of Git changesets, the fully-automated analysis cost is viable, with an average running time of \AvgRuntimePerLog\ secs per logging statement and \AvgRuntimePerKLOCChanged\ secs per thousand lines of code changed,

	\item developers do not actively think about how their logging statement levels evolve with their software,
		motivating an automated approach, and

	\item our approach is promising in evolving feature log levels.
\end{enumerate*}


This work's contributions are summarized as follows: 

\begin{description}

	\item[Approach design.] We present an automated approach that programmatically manipulates a Mylyn DOI model using Git histories to evolve feature logging statement levels to better align with the current features of interest.
	    Widespread manual log level modification is alleviated, information overload is reduced, and more relevant events are underscored, potentially exposing bugs.

	\item[Heuristic formulation.] Heuristics---based on first-hand developer feedback---to distinguish between feature logs and those with more critical information are proposed.

	\item[Implementation \& experimental evaluation.] To ensure real-world applicability, we implemented our approach as an open-source Eclipse IDE plug-in built upon Mylyn and JGit and used it to study \NumProjects\ Java projects.
	    Our technique successfully analyzes \analyzedPercentage\ of
	    logging statements, increases log level distributions by
	    \distributionPercentageIncreaseApprox, 
		and increases the focus of logs in bug fix contexts at a rate of \bugsIdealVersusActualRateApprox. Furthermore, several pull requests were integrated
		into
		large and popular open-source projects.

\end{description}



\section{Motivating Example}\label{sec:motiv}

\begin{listing}[t]
	\begin{javacode}
		public class Wombat {|\label{lne:wombat}|
			private static final Logger logger = Logger.getLogger("global");|\label{lne:logger}|
			logger.setlevel(Level.FINE);// Only logs |$\geq$| FINE.
			private double temp; private double oldTemp;|\label{lne:temp}|

			public void setTemp(double val) {|\label{lne:setTempStart}|
				this.oldTemp = temp;|\label{lne:setTemp}| this.temp = val;|\label{lne:cache}|
				logger.log(Level.FINER, "Temp set to: " + this.temp);|\label{lne:logTemp}|
				logger.finer("Old temperature was: " + this.oldTemp);|\label{lne:logOldTemp}|}

			public static void main(String[] args) {|\label{lne:mainStart}|
				Wombat w = new Wombat();|\label{lne:createWombat}|
				Scanner scanner = new Scanner(System.in);

				System.out.println("Enter a temperature:");
				double input = scanner.nextDouble();|\label{lne:ask}| w.setTemp(input);|\label{lne:tempSet}|

				try { // send to file.|\label{lne:fileStart}|
					logger.fine("Writing to file.");|\label{lne:fileLog}|
					Files.writeString("output.txt", w.toString(), WRITE);|\label{lne:write}|
				} catch (IOException e) { // Fatal error.
					logger.severe("Couldn't open file for writing.");|\label{lne:logError}|
					throw e;}|\label{lne:fileEnd}|}}
	\end{javacode}
	\caption{Hypothetical logging usage example~\cite{QOS.ch2019a}.}\label{lst:wombat}
\end{listing}

\Cref{lst:wombat} portrays a hypothetical code snippet~\cite{QOS.ch2019a} that uses \javainline{java.util.logging} (JUL)~\cite{Oracle2018b} having log levels that include---in ascending order---\ji{FINEST}, \ji{FINER}, \ji{FINE}, \ji{INFO}, \ji{WARNING}, and \ji{SEVERE}. A \javainline{Wombat} class starts at line~\ref{lne:wombat} and has a \javainline{logger} (line~\ref{lne:logger}) and current and previous \javainline{temp}eratures (line~\ref{lne:temp}). The \javainline{logger} is configured so that only logs with levels $\geq$ \javainline{FINE} are emitted to reduce information overload.

A mutator for \javainline{temp} begins on line~\ref{lne:setTempStart}. On line~\ref{lne:setTemp}, old \javainline{temp} values are cached. Then, new and old temperatures are logged on lines~\ref{lne:logTemp} and~\ref{lne:logOldTemp}, respectively.
Both statements log at the \javainline{FINER} level.
Since \javainline{logger} has been previously configured not to emit logs with levels $\leq$ \ji{FINER}, the statements have no effect.

When creating \javainline{Wombat}s (line~\ref{lne:createWombat}), the user is asked for a temperature (line~\ref{lne:ask}). A string representation of the \javainline{Wombat} not shown in this paper is then saved to a file (lines~\ref{lne:fileStart}--\ref{lne:fileEnd}). Line~\ref{lne:fileLog} logs that the writing has commenced, and since the level is \ji{FINE},
the statement
emits a log. The actual file writing takes place on line~\ref{lne:write}. Because \javainline{Files.writeString()} possibly throws an \javainline{IOException}, the call is surrounded by a try/catch block. Line~\ref{lne:logError} executes when the specified exception has been caught. This log message is emitted since \ji{SEVERE} $\geq$ \ji{FINE}.

\begin{listing}[t]
	\begin{diffcode}
@@ -23,11 +23,15 @@ public void setTemp(double val) {
+ if (val > 0) {
|\label{lne:guardStart}|    this.oldTemp = temp; this.temp = val;
|\label{lne:finer1}|		logger.log(Level.FINER, "Temp set to: " + this.temp);
|\label{lne:guardEnd}|    logger.finer("Old temperature was: " + this.oldTemp);}
|\label{lne:except}|+ else throw new IllegalArgumentException("Invalid: " + val);

@@ -38,7 +42,17 @@ public static void main(String[] args)
|\label{lne:clientStart}|+ while (true) {
+   try {
      w.setTemp(input);
+     break; // succeeded.
+   } catch (IllegalArgumentException e) {
+     // Not a fatal error. Log the exception and retry.
|\label{lne:caught}|+     logger.log(Level.INFO, "Invalid input: " + input, e);
|\label{lne:clientEnd}|+     System.out.println("Invalid temp. Please retry.");}}
	\end{diffcode}
	\caption{Rejecting invalid temperatures.}\label{lst:wombat2}
\end{listing}

\begin{listing}[t]
	\begin{diffcode}
@@ -30,6 +30,9 @@ public void setTemp(double val) {
  else throw new IllegalArgumentException("Invalid:" + val);
+
+ if ((this.temp - this.oldTemp) / this.oldTemp > 0.05)
|\label{lne:degrees}|+   logger.warning("Temperature has risen above 5%.");
	\end{diffcode}
	\caption{Warning about drastic temperature changes.}\label{lst:wombat3}
\end{listing}

\Cref{lst:wombat2} depicts a changeset\footnote{Although only additions are shown, similar issues may arise with deletions.}
where invalid \ji{temp}eratures (i.e., negative temperatures) are rejected by guarding lines~\ref{lne:guardStart}--\ref{lne:guardEnd} and throwing an exception on line~\ref{lne:except}. As a result, client code (lines~\ref{lne:clientStart}--\ref{lne:clientEnd}) is modified to handle the exception, looping until valid input is entered. On line~\ref{lne:caught}, a log is issued when the exception is caught, documenting the retry. Because the error is non-fatal, \ji{INFO} is used. An ensuing changeset (\cref{lst:wombat3}) logs a warning (line~\ref{lne:degrees}) when temperatures increase by more than \SI{5}{\percent}.

\begin{listing}[t]
	\begin{javacode*}{firstnumber=5}
		public void setTemp(double val) {
			if (val > 0) {
				this.oldTemp = temp; this.temp = val;
				logger.log(Level.FINE|\st{R}|, "Temp set to: " + this.temp);|\label{lne:transformFine}|
				logger.fine|\st{r}|("Old temperature was: " + this.oldTemp);|\label{lne:transformFine2}|}
			else throw new IllegalArgumentException("Invalid:" + val);

			if ((this.temp - this.oldTemp) / this.oldTemp > 0.05)
				logger.warning("Temperature has risen above 5%.");|\label{lne:tempWarn}|}

		public static void main(String[] args) { // ...
				} catch (IllegalArgumentException e) {
					// Not a fatal error. Log the exception and retry.
					logger.log(Level.INFO, "Invalid input: "+ input, e);|\label{lne:invalidInput}|
					System.out.println("Invalid temp. Please retry.");}}

			try { // send to file.
				logger.fine|\smash{\ul{st}}|("Writing to file.");|\label{lne:fileDecrease}|
				Files.writeString("output.txt", w.toString(), WRITE);
			} catch (IOException e) { // Fatal error.
				logger.severe("Couldn't open file for writing.");|\label{lne:fileError}| // ...
	\end{javacode*}
	\caption{Resulting ``reinvigorated'' logging levels.}\label{lst:wombat4}
\end{listing}


\Cref{lst:wombat4} shows an abbreviated result, containing only relevant parts of \cref{lst:wombat} with \cref{lst:wombat2,lst:wombat3} applied and transformations made to feature logging statement levels. Recent changes
to nearby code of feature logging statements may indicate that features, e.g., temperature management, implemented at this part of the code are of a higher ``interest.''
As such, the levels at lines~\ref{lne:transformFine}--\ref{lne:transformFine2} have \emph{increased} from \ji{FINER} to \ji{FINE}, potentially
helping developers debug the new feature code.
In this example, the transformed logs will now emit.

Contrarily, as file writing is \emph{not} being actively developed---the feature logging statement level at line~\ref{lne:fileDecrease} \emph{decreased} from \ji{FINE} to \ji{FINEST}. Recent changesets did \emph{not} include edits to this region, thus resulting in log suppression. While code recently edited may have regressions, the (non-feature) logging statement
at line~\ref{lne:fileError} did \emph{not} have its level lowered and remains useful in finding possible regressions. Likewise, the non-feature logging statement at line~\ref{lne:invalidInput}, although non-fatal, was \emph{not} lowered \emph{despite} recent non-local edits.

As logging is pervasive~\cite{Zeng2019}, manually managing feature logging statement levels
can be overwhelming. Even this simple example demonstrates that logging statements can be scattered and tangled with code implementing core functionality. Automatically evincing information related to features that are developed more and more often to the forefront,
while gradually suppressing those less and less frequently developed may enhance focus and help expose potential bugs.

\section{Approach}\label{sec:approach}



\subsection{Overview}

\begin{figure}
    \centering
    \includegraphics[width=\linewidth,keepaspectratio]{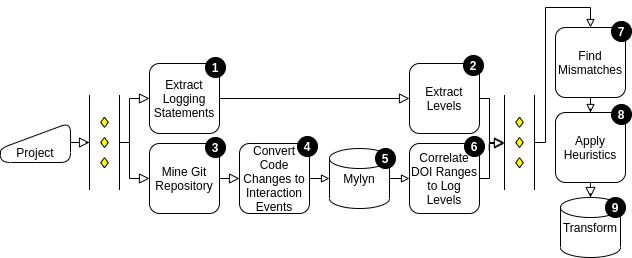}
    \caption{Logging level revitalization approach overview.}\label{fig:overview}
\end{figure}


Our automated approach
(\cref{fig:overview})
programmatically manipulates a Mylyn DOI model~\cite{Kersten2005} by mining Git repositories to evolve feature logging statement levels to better align with the current features of interest.
Mylyn~\cite{Kersten2006} has traditionally been used to associate IDE edit and graphical interaction events to a particular \emph{task} to support task \emph{context switching}. The two branches between yellow diamonds could be executed concurrently. We adapt this model to track the ``interest'' of code elements surrounding feature logging statements (step 4) to find \emph{mismatches} (step 7) between log levels and feature interests for feature logging statements. Furthermore, to aggregate interest levels across a development team, edits are extracted from Git histories (step 3) instead of IDE interactions. The DOI model is then programmatically manipulated using the extracted edits (step 5) and finally partitioned (step 6) and compared to the current feature logging statement (step 1) levels (step 2). If a mismatch is found, the level is transformed in the code (step 9). To distinguish feature logging statements from other logging that could be more critical and to guide the transformation, a set of heuristics that we define are used (step 8).

\subsection{Feature Logging Statement Level Extraction}

Logging statements---later used for
\begin{enumerate*}[(i)]
    \item correlating the degree of interest of surrounding code when finding mismatches (step 7),
    \item applying heuristics (step 8), and
    \item as potential transformation sites (step 9)---are
\end{enumerate*}
extracted in step 1. Current levels are extracted from the statements (step 2) and later used in mismatch identification (step 7). Depending on the API used, level extraction is performed in two ways. For instance, JUL has both convenience and standard logging APIs. Convenience APIs have method names that correspond to the desired log level (e.g., line~\ref{lne:logOldTemp}, \cref{lst:wombat}). On the other hand, standard APIs have logging levels passed as parameters (e.g., line~\ref{lne:logTemp}, \cref{lst:wombat}). In both cases, Abstract Syntax Trees (ASTs) are extracted from the underlying source code. Whereas the convenience case is straight-forward, in the standard case, our current implementation only extracts levels represented as literals. Using data-flow analysis
is a subject of future work; however, we analyzed \analyzedPercentage\ of logging statements during our study successfully despite this limitation.

\subsection{Mylyn DOI Model Manipulation}\label{sec:mylyn}

\subsubsection{Background}\label{sec:mylyn:background}

Mylyn~\cite{EclipseFoundation2020} maintains focused contexts of entities relevant to a particular task using a DOI model. A context comprises the relevant elements (e.g., classes, methods, fields), along with how \emph{interesting} the elements are to the related task. The more a developer \emph{interacts} with an element (e.g., navigates to a file, edits a file) when working on a task, the more interesting the element is deemed to be, and vice-versa. Elements also \emph{decay}, i.e., as other elements increase in DOI, elements that were once ``interesting'' decrease. Mylyn then alters the IDE's behavior so that interesting elements are displayed more prominently throughout its views.


\subsubsection{Repository Mining}

The Mylyn context is adapted to ascertain the interest levels of code surrounding logging statements. Traditionally, Mylyn is used for \emph{context switching}; i.e., relevant elements are stored in task contexts. That way, developers can easily switch between tasks without losing their focus on related code elements. It is confined to a single developer's workspace; however, code modifications made by our approach are \emph{global}---affecting all project developers. As such, the context is ``expanded'' to include \emph{all} developers' interests by mining Git repositories in step 3.

Mylyn can record a wide range of interaction events between developers and program elements, such as element selection, element editing, etc. However, our approach only takes into account the interaction event of element editing because (1) edits can be easily saved, for example, through Git histories (2) edits can more directly express developers' interest than other interaction events, such as element selection.

\subsubsection{Converting Code Changes to Interaction Events}\label{sec:convert}

A central notion of Mylyn are ``interaction events,'' which dictate how the DOI model is manipulated. The more interaction a particular element has, the larger its DOI and vice-versa. Although Mylyn has a broad spectrum of interaction event types, we focus on ``edit'' events as we mine Git code changesets. 

In converting Git code changes to Mylyn interaction events
(step 4), we mainly focus on changes to method and constructor bodies. While edits to other kinds of code elements, e.g., fields, could be near logging statements, this is currently not supported and is not a representative use case. Furthermore, Git edits are parsed
instead of AST differencing, which can be computationally expensive. Moreover, AST differencing does not include whitespace changes, which are desirable
as they may indicate interest.\footnote{We also consider changes to non-source lines, e.g., comments.}
Interaction events are then processed by Mylyn as if they had emanated from the IDE in step 5.

\paragraph{Rename Refactorings \& Copying}

Program elements (e.g., methods) altered in Git repositories may no longer exist in the current version where level transformations would take place. Such historical elements that were removed are ignored as they do not exist in the current version. However, elements that have undergone rename refactorings need to be considered as they will have a counterpart in the current version.
To this end, during repository mining, we maintain a data structure that associates rename relationships between program elements, e.g., method signatures. Before converting changesets to interaction events, a lookup is first performed
to retrieve the changed element's signature in the current version. Unfortunately, handling refactorings necessitates two history traversals, one to create the renaming data structure and the other to process changesets. However, in our implementation, we have a performance improvement where code changes are cached during the renamings detection, and only code changes are traversed subsequently rather than the entire Git history.



Because only rename refactorings are needed, instead of more advanced approaches~\cite{Tsantalis2018}, our current implementation uses lightweight
approximations, such as basic method signature and body similarity comparison. Nevertheless, during our evaluation, we were able to successfully analyze the change history of code surrounding \analyzedPercentage\ of \logs\ logging statements across \NumProjects\ projects. In the future, we will explore integrating more advanced techniques.

For copying, we use the copy detection features of Git at the file level. If Git detects a copied file, any DOI values associated with the original file serve as the starting values for elements in the new file. We will explore integrating more advanced copy detection, e.g., for methods~\cite{Chang2008}, in the future.

\subsection{DOI-Feature Logging Level Association}

Step 5 results in a rich
DOI model where the most and most recently edited code is correlated with the highest DOI and vice-versa. Final DOI values that are non-negative reals
are then partitioned so that DOI ranges can be associated with log levels
(step 6). The association is then used to discover mismatches between interest and feature logging statement levels (step 7).


Partitions are created by subtracting the largest DOI value
by the smallest and dividing the result by the number of available levels,
producing a DOI partition range size. Then, each DOI range is associated with levels by order. For example, the least logging level (e.g., \ji{FINEST}) is associated with the first DOI partition (e.g., $[0,2.54)$). However, this scheme can be influenced by the log category heuristic, i.e., treating \ji{WARNING} and \ji{SEVERE} as categories rather than levels. In such cases, specific partitions will dynamically not cause mismatches to be detected, potentially affecting the transformations performed. We utilize equal range size of DOI partitions as the initial step of research because it is the most straightforward way to deal with the matching. In the future, we will consider Machine Learning approaches to improve our algorithms.

The above scheme creates equivalently-sized partitions. Elements tend not to change often over time; therefore, a na\"{\i}ve partitioning may result in uneven distribution. Luckily, Mylyn supports customizable element decay rates (cf.~\cref{sec:mylyn:background}), where---as other elements become more interesting---less and less frequently edited elements lose their ``interest'' at a specific rate. The default decay rate does not suffice because Mylyn was not originally designed to process the sheer number of modifications typical found in Git repositories contributed by multiple developers working on many tasks. Instead, it was designed to record \emph{IDE interactions} made by a \emph{single} developer working on a \emph{single} task. Hence, the default decay rate causes elements to decrease in DOI rapidly; thus, we decreased the rate significantly. The decay rate was calculated by analyzing two open-source projects (i.e., \selenium\ and \IRCT). It was later tweaked to fit into all of the projects we studied. Although several partitioning schemes
were attempted, we
found that combining equivalently-sized partitions with a reduced decay rate worked the best.

Once partitions are formed, potential mismatches are discovered (step 7) by comparing partitions with the current
statement levels from step 2. 
If a mismatch is found, the statement is marked for potential transformation barring heuristics.

%

\subsection{Feature Logging Statement Classification \& Heuristics}\label{sec:heuristics}

Distinguishing between feature logging statements and other kinds of logging---and transformation guidance---is accomplished via heuristics (step 8). 
We conducted a pilot study to assist in the design of heuristics through trial and error. We used an early version of the tool and sent pull requests to large and popular open-source projects with a small subset of the tool's output. After that, we obtained feedback from the developers. We investigated the code surrounding logging statements based on their feedback to improve our approach.
Heuristics are used to avoid undesirably transforming logging statements, e.g., lines~\ref{lne:tempWarn},~\ref{lne:invalidInput}, and~\ref{lne:fileError} of \cref{lst:wombat4}. Here is a list of heuristics, with their abbreviations in bold font:

\begin{enumerate}

	\item \textbf{WS}: Treat particular levels as log \emph{categories}.\label{itm:CWS}

	\item \textbf{LOW}: Never \emph{lower} the level of logging statements:\label{itm:lower}

	    \begin{enumerate}

		\item \textbf{CTCH}: appearing within catch blocks.\label{itm:catch}

		\item \textbf{IFS}: immediately following branches (e.g., if, else, switch).\label{itm:followingifs}

		\item \textbf{KEYL}: \emph{having} particular keywords in their log messages.\label{itm:withkeys}

	    \end{enumerate}

	\item \textbf{CNDS}: Never \emph{change} the level of logging statements immediately following branches whose condition contains a log level.\label{itm:wrapping}

	\item \textbf{KEYR}: Never \emph{raise} the level of logging statements \emph{without} particular keywords in their log messages. This only applies for target levels \ji{WARNING} or \ji{SEVERE} (\ji{WARN} and \ji{ERROR} in SLF4J) ---typically used in more critical situations---and does not apply when \hyperref[itm:CWS]{\textbf{WS}} is enabled.\label{itm:withoutkeys}

	\item \textbf{INH}: Only \emph{consistently} transform the level of logging statements appearing in overriding methods.\label{itm:subtyping}

	\item \textbf{TDIST}: Only transform the level of logging statements up to a transformation distance threshold.\label{itm:threshold}

\end{enumerate}

\paragraph{Logging Categories}

For \hyperref[itm:CWS]{\textbf{WS}}, a developer may choose to treat \ji{WARNING} and \ji{SEVERE}
as logging statement categories rather than traditional levels. Such log levels are more likely to be associated with failures or potential bugs rather than the feature implementation surrounding the logs. This way, developers can denote that logging statements with such levels have special semantics and are not on a ``sliding scale.'' Denoting \ji{WARNING} and \ji{SEVERE} as categories can be used to avoid transforming lines~\ref{lne:tempWarn} and~\ref{lne:fileError} of \cref{lst:wombat4}. In SLF4J, developers could use level \ji{WARN} and \ji{ERROR} as logging statement categories.




\paragraph{Catch Blocks \& Branches}

Logging statements appearing in catch blocks may serve as error notifications~\cite{OracleCatch}. This critical logging information for bug detection in catch blocks should not be overlooked, so we proposed~\hyperref[itm:catch]{\textbf{CTCH}} that ensures that the level of these statements is never reduced. For example, line~\ref{lne:fileError}, \cref{lst:wombat4} did not have its level lowered despite the lowering of a level at line~\ref{lne:fileDecrease} due to recent non-local edits.
\hyperref[itm:followingifs]{\textbf{IFS}} is similar to \hyperref[itm:catch]{\textbf{CTCH}}, with an example at line~\ref{lne:tempWarn}, \cref{lst:wombat4}.
Below is an abbreviated example from \blueocean~\cite{McDonald2018}, where two similar logging statements appear in each of the different contexts:

\begin{javacode*}{linenos=false}
	try {
		node = execution.getNode(action.getUpstreamNodeId());
	} catch (IOException e) {
		LOGGER.warning("Couldn't retrieve upstream node: " + e);}
	if (node == null) {
		LOGGER.warning("Couldn't retrieve upstream node (null)");}
\end{javacode*}

\paragraph{Log Wrapping}

\hyperref[itm:wrapping]{\textbf{CNDS}} prevents logging semantic violations when run-time level checks redundantly guard logging statements. Consider the below example from \guava~\cite{Decker2014}:

\begin{javacode}
	if (logger.isLoggable(Level.SEVERE))|\Vlabel{lne:guard}|
		logger.log(Level.SEVERE, message(ctxt), exception);|\Vlabel{lne:guardlevel}|
\end{javacode}

\noindent Altering the level at \lref{lne:guardlevel} without also changing the level at \lref{lne:guard} would be counterproductive. More sophisticated analysis is necessary to handle such cases, which are not typical. Moreover, other approaches~\cite{Li2017a} do not deal with this.

\paragraph{Keywords}

\hyperref[itm:withkeys]{\textbf{KEYL}} and \hyperref[itm:withoutkeys]{\textbf{KEYR}} help distinguish feature logging statements using keywords, which originated from our evaluation and developer feedback. For the former, we manually assessed the transformations made by earlier versions of our approach, comparing them to the surrounding context. For the latter, developers commented on the transformations made by earlier versions of our tool. In both cases, we noted common keywords that appeared in logging statement messages. These words elicit strong emotions. Rather than the feature implementation surrounding the logs, they are more likely to have an association with probable issues.
Stopgap words are used to maximize coverage, and keywords must appear in the literal parts of the log message construction. Keywords for \hyperref[itm:withkeys]{\textbf{KEYL}} include ``fail,'' ``disabl,'' ``error,'' and ``exception;'' keywords for \hyperref[itm:withoutkeys]{\textbf{KEYR}} include ``stop,'' ``shut,'' ``kill,'' ``dead,'' and ``not alive.'' 

For example, in \cref{lst:wombat4}, the levels at lines~\ref{lne:transformFine}--\ref{lne:transformFine2} are allowed to \emph{increase} from \ji{FINER} to \ji{FINE} because the target levels are neither \ji{WARNING} nor \ji{SEVERE}, passing \hyperref[itm:withoutkeys]{\textbf{KEYR}}. The level at line~\ref{lne:fileDecrease} was allowed to \emph{decrease} from \ji{FINE} to \ji{FINEST} as there are no ``anti-lowering'' keywords (\hyperref[itm:withkeys]{\textbf{KEYL}}). In contrast, the levels at lines~\ref{lne:invalidInput} without local edits and~\ref{lne:fileError} were not lowered, partly due to having anti-lowering keywords.

While the keywords are not exhaustive, they were derived via an extensive study and assistance of open-source developers.
Nevertheless---in the future---we will explore using machine learning (ML) for broader classification,
as well as adding more keywords
related to security and privacy.

%
%
%
%

\paragraph{Subtyping}

\hyperref[itm:subtyping]{\textbf{INH}}---formulated using developer feedback---applies when mismatches are found in methods involved with inheritance.
Specifically, if method $M'$ overrides method $M$ and both $M$ and $M'$ include level mismatches, the
target levels
must be consistent as to preserve a behavioral subtyping-like~\cite{Leavens1990,Liskov1994} relationship w.r.t.~logging.

\paragraph{Transformation Distance}

Because logging is pervasive, our approach may suggest widespread modifications. As such, \hyperref[itm:threshold]{\textbf{TDIST}}---also from developer feedback---curtails the degree of level transformations using a threshold, which is a setting in our tool. Finally, all mismatches that pass the enabled heuristics are transformed via AST node replacements (step 9).




\section{Evaluation}\label{sec:eval}

\subsection{Implementation}\label{sec:impl}

Our approach is implemented as an
open-source Eclipse IDE~\cite{EclipseFoundation2020a} plug-in~\cite{Tang2020} and built upon JGit~\cite{EclipseFoundation2020b}, for Git
extraction, and Mylyn~\cite{EclipseFoundation2020}, for
DOI
manipulation. Eclipse is leveraged for its extensive source-to-source transformation support~\cite{Baeumer2001}
and that it is entirely open-source for all Java development.
It also supports widely-used build systems, so that projects using different IDEs may also use our plug-in.

Mylyn provides many DOI facilities,
including model tuning, interaction event modeling, and interaction prediction. Integrating with Mylyn also makes it possible to---in the future---combine historical edit events from Git with IDE interactions---potentially leading to a richer DOI model.
It may also be possible to extend Mylyn to populate DOI models with version control events~\cite{Pilgrim2014},
solving an open Eclipse bug~\cite{Davis2014}.

Eclipse ASTs with source symbol bindings are used as an intermediate
representation. Two popular logging frameworks, namely, JUL~\cite{Oracle2018b} and SLF4J~\cite{QOS.ch2019}, are currently supported.
Heuristics (\cref{sec:heuristics}) are presented as tool options.

Although we
depend on Mylyn for
DOI
and Eclipse for code analysis, transformation, and preview-panes, it may be possible to convert our tool to a GitHub App~\cite{GitHub2020} that would monitor Git commits, periodically update an isolated DOI model, and generate recommended modifications
as pull requests.
This future work may follow
recent work on refactoring bots~\cite{Alizadeh2019}.

\subsection{Experimental Evaluation}

\subsubsection{Research Questions}\label{sec:eval:questions}


We answer the following questions:

\begin{questions}

    \item How applicable is our tool to and how does it behave with real-world open-source software?\label{rq:appl}

















    \item Can our tool help bring focus to buggy code?\label{rq:bugs}

    \item
	Are our tool's results acceptable? What is its impact?\label{rq:useful}

\end{questions}

\ref{rq:appl}~answers whether the proposed approach scales, in terms of SLOC, number and usages of logging statements, and revision history length, to real-world projects. It also provides insight on how logging statements are used and the contexts for which they appear by 
\begin{enumerate*}[(i)]
    \item assessing heuristics applicability,
    \item studying the degree and directionality of mismatch between edit frequency and logging levels, and
    \item measuring level distribution before and after our tool's application for fuller logging level spectrum usage by developers.
\end{enumerate*}

\ref{rq:bugs} inquires about our tool's ability to increase buggy code focus by altering feature logging statement levels. To help potentially expose bugs, \ref{rq:bugs} assesses whether our tool increases levels of feature logging statements in the context of bug fixes (i.e., ``buggy'' contexts) and likewise decreases levels in non-``buggy'' contexts. Lastly, \ref{rq:useful} gauges whether the transformations are acceptable and their impact on developer communities. 
The former effectively evaluates
the heuristics,
while the latter assesses the extent to which our tool's transformations affect developers as a whole.

To answer~\ref{rq:appl}, quantitative (\cref{sec:eval:quant}) and qualitative (\cref{sec:eval:qual}) analyses are performed.
To answer~\ref{rq:bugs}, we applied our tool to software versions leading up to buggy and non-buggy feature log contexts
mined from software repositories. Finally, a pull request study is issued to answer~\ref{rq:useful} (\cref{sec:eval:pulls}). Our dataset~\cite{Anonymous2020a} is available.

\subsubsection{Quantitative Analysis}\label{sec:eval:quant}

We ran our tool on a large corpus.

\begin{sidewaystable}
		\resizebox{\textwidth}{!}{%
			\begin{threeparttable}
				\begin{tabular}{lllllllllllllllllllll}
					\toprule
					subject & HEAD & KLOC & Kcms & $\delta$KLOC & fw & logs & fails & trns & ctch & ifs & cnds & keyl & keyr & inh & $\overline{\mathit{dist}}$ ($\sigma_{\mathit{dist}}$) & $\sigma_{\mathit{pre}}$ & $\sigma_{\mathit{post}}$ & low & rse & t (m) \\
					\midrule
					bc-java & a66e904 & 701.57 & 6.00 & 1,213 & 1 & 56 & 4 & 11 & 5 & 13 & 6 & 16 & 0 & 0 & 2.64 (1.30) & 1.49 & 2.01 & 11 & 0 & 62.78 \\
					blueocean\tnote{*} & c3bfac5 & 49.32 & 4.06 & 138 & 2 & 109 & 0 & 13 & 11 & 6 & 5 & 57 & 9 & 0 & 1.54 (0.63) & 1.65 & 1.74 & 11 & 2 & 2.93 \\
					californium & 5b026ab & 79.35 & 1.00 & 144 & 1 & 986 & 0 & 141 & 23 & 123 & 95 & 317 & 176 & 2 & 1.52 (0.73) & N/A & N/A & 118 & 22 & 7.51 \\
					CoreNLP & d7356db & 535.07 & 6.00 & 43,686 & 2 & 358 & 0 & 146 & 0 & 87 & 21 & 55 & 0 & 0 & 3.10 (1.43) & 1.84 & 2.16 & 146 & 0 & 444.99 \\
					deviation\tnote{*} & 88751d6 & 6.52 & 0.08 & 25 & 1 & 91 & 0 & 11 & 1 & 7 & 8 & 48 & 3 & 0 & 2.64 (1.23) & 1.69 & 2.07 & 11 & 0 & 0.06 \\
					errorprone & 2118ba3 & 164.65 & 3.99 & 517 & 1 & 12 & 0 & 1 & 1 & 0 & 1 & 9 & 0 & 0 & 2.00 (0.00) & 1.61 & 1.86 & 1 & 0 & 17.36 \\
					guava & 71de406 & 393.69 & 1.00 & 297 & 1 & 36 & 0 & 0 & 0 & 0 & 0 & 33 & 0 & 0 & 0.00 (N/A) & 1.52 & 1.52 & 0 & 0 & 29.63 \\
					hollow & ff635ee & 68.60 & 0.90 & 159 & 1 & 31 & 0 & 6 & 0 & 3 & 0 & 22 & 0 & 0 & 3.50 (1.12) & 1.02 & 1.98 & 5 & 1 & 2.86 \\
					IRCT & d67f539 & 42.29 & 0.89 & 194 & 1 & 13 & 0 & 6 & 0 & 0 & 0 & 2 & 1 & 0 & 2.00 (1.41) & 1.22 & 1.51 & 6 & 0 & 2.07 \\
					JacpFX & 14c2a4c & 24.06 & 0.37 & 121 & 1 & 21 & 0 & 2 & 4 & 0 & 7 & 8 & 0 & 0 & 2.50 (1.50) & 1.47 & 1.69 & 2 & 0 & 0.77 \\
					jdp\tnote{*} & 515b7e7 & 35.86 & 1.00 &	100 & 2 & 799 & 0 & 180 & 17 &	15 & 4 & 84 & 399 &	0 &	1.70 (0.63) & 0.00 & 1.79 & 139 & 41 & 3.18 \\
					jenkins & 4f0436a & 160.60 & 1.00 & 25 & 1 & 503 & 22 & 60 & 29 & 52 & 54 & 244 & 22 & 0 & 2.17 (1.25) & 1.73 & 1.84 & 58 & 2 & 3.81 \\
					jsilhouette & 8de4e64 & 2.35 & 0.03 & 6 & 1 & 22 & 0 & 0 & 0 & 8 & 0 & 14 & 0 & 0 & 0.00 (N/A) & 0.00 & 0.00 & 0 & 0 & 0.04 \\
					junit5 & 3969585 & 73.59 & 5.98 & 381 & 1 & 9 & 2 & 4 & 0 & 0 & 0 & 2 & 0 & 0 & 3.00 (1.00) & 0.00 & 1.67 & 4 & 0 & 75.25 \\
					OpenGrok & 0f3b14d & 71.65 & 4.93 & 377 & 1 & 592 & 3 & 61 & 44 & 49 & 41 & 353 & 13 & 0 & 2.70 (1.47) & 1.81 & 2.07 & 53 & 8 & 15.20 \\
					selenium & a49fb60 & 93.61 & 1.00 & 76 & 1 & 94 & 0 & 28 & 5 & 6 & 0 & 48 & 4 & 0 & 2.96 (0.94) & 1.00 & 1.75 & 27 & 1 & 1.45 \\
					SpotBugs & ecc7994 & 187.78 & 1.00 & 94 & 2 & 96 & 0 & 44 & 2 & 15 & 6 & 26 & 0 & 0 & 1.57 (0.96) & 1.74 & 2.30 & 44 & 0 & 26.59 \\
					WALA & b73a0b2 & 202.45 & 1.00 & 236 & 2 & 145 & 0 & 39 & 1 & 38 & 0 & 22 & 16 & 1 & 1.41 (0.54) & 0.94 & 0.94 & 38 & 1 & 9.20 \\
					Total &  & 2,893.01 & 40.22 & 47,787 & 1.28 & 3,973 & 31 & 753 & 143 & 422 & 248 & 1,360 & 643 & 3 & 2.13 (1.23)\tnote{\textdagger} & 1.75 & 2.05 & 674 & 78 & 705.68 \\
					\bottomrule
				\end{tabular}%
				\begin{tablenotes}
					\item[*] blueocean is \blueocean, deviation is \deviation, and jdp is \jdp.
					\item[\textdagger] Overall average and standard deviation.
				\end{tablenotes}
			\end{threeparttable}
		}
	\caption{Quantitative analysis results.
		Column \textbf{Kcms} is the thousands of commits analyzed.
		Column \textbf{$\delta$KLOC} is the thousands of source lines of code.
		Column \textbf{fw} is the number of logging frameworks detected.
		Column \textbf{logs} is the sum of candidate logging statements.
		Column \textbf{fails} is the number of statements where the current level could not be extracted.
		Column \textbf{trns} is the number of transformed logging statements.	
		Column \textbf{ctch} is the sum of log levels not lowered due to \hyperref[itm:catch]{\textbf{CTCH}}.		
		Column \textbf{ifs} is the sum of log levels not lowered due to \hyperref[itm:followingifs]{\textbf{IFS}}.
		Column \textbf{keyl} is the sum of log levels not lowered due to \hyperref[itm:withkeys]{\textbf{KEYL}}.
		Column \textbf{cnds} is the sum of log levels not transformed due to \hyperref[itm:wrapping]{\textbf{CNDS}}.
		Column \textbf{keyr} is the sum of log levels not raised due to \hyperref[itm:withoutkeys]{\textbf{KEYR}}.	
		Column \textbf{inh} is the sum of levels not modified due to \hyperref[itm:subtyping]{\textbf{INH}}.
		Column \textbf{$\overline{\mathit{dist}}$} is the average level transformation distance and corresponding standard deviation ($\sigma_{\mathit{dist}}$).	
		Columns \textbf{$\sigma_{\mathit{pre}}$} and \textbf{$\sigma_{\mathit{post}}$} are the stdev of log levels \emph{before} and \emph{after} transformation for JUL, respectively.		
		Column \textbf{lower} is the number of lowered log levels.
		Column \textbf{raise} is the number of raised log levels.			
		Column \textbf{t (m)} is the total running time in minutes.
	}\label{tab:results}
\end{sidewaystable}

\paragraph{Subject Selection \& Details}

This phase involves \NumProjects\ open-source Java applications and libraries of varying sizes and domains (\cref{tab:results}). Subjects were also chosen to include logging statements using either JUL or SLF4J and have at least one source code change to a method containing a logging statement (i.e., a candidate statement).
Column \textbf{KLOC} ranges from \num{\sim6}K for \deviation\ to \num{\sim535}K for \CoreNLP\@. Column \textbf{Kcms} can significantly affect the number of candidate statements.
For subjects with long Git histories,
we choose a large number of commits
to obtain more candidates. For others, a relatively small number of commits was adequate to obtain all logging statements as candidates.

\paragraph{Execution \& Run Time}

The analysis was executed on an Intel Xeon-E3 with four cores, 31 GB RAM, and a 20 GB maximum heap size. Column \textbf{t (m)} averages  \AvgRuntimePerLog\ secs per candidate statement and \AvgRuntimePerKLOCChanged\ secs per KLOC changed ($\delta$KLOC). The running time is highly related to analyzed source lines per commit (\textbf{$\delta$KLOC}/Kcms),
with a \num{0.98} Pearson correlation coefficient that ranges from \num{-1} to \num{1} and \num{1} is an exact linear correlation.
\CoreNLP, having a particularly long and intricate Git history, is an outlier, taking over half of the running time.
\takeaway{Logging statements process in \AvgRuntimePerLog\ secs and one $\delta$KLOC in \AvgRuntimePerKLOCChanged\ secs, with the processing time highly connected to LOC/commit (\num{0.98} Pearson correlation).}

\paragraph{Log Level Reinvigoration}

We successfully analyzed \analyzedPercentage\ of \logs\ candidate statements (column \textbf{logs}) across \NumProjects\ subjects.\footnote{\junit\ only has \num{9} candidates as this version mainly uses custom logging. It is included since its previous versions use JUL and are subsequently studied.}
 Failures in column \textbf{fails} include when the logging level is stored in a variable.\takeaway{\analyzedPercentage\ of \logs\ logging statements were successfully analyzed.}


Column \textbf{trns} denotes \transPercent\ of candidate logging statements are transformed.
This metric is not a success rate as it is dependent on the mismatches found and the enabled heuristics. Instead,
it demonstrates that the transformations made by our tool---when applied to real-world projects---are subtle enough that they may be appealing to more risk-averse projects.
It also
shows that
manual effort can be labor-intensive, involving multi-developer, historical analysis and transformation, as manually modifying \trans\ logging statements is non-trivial.\takeaway{The transformations were subtle (\transPercent) yet labor-alleviating (\trans\ transformations).}

\paragraph{Heuristics} 

To more fully understand the effects, all heuristics were enabled except log categories and transformation distance (\hyperref[itm:CWS]{\textbf{WS}} and \hyperref[itm:threshold]{\textbf{TDIST}}, respectively, in \cref{sec:heuristics}) in this phase. Columns \textbf{ctch}, \textbf{ifs}, and \textbf{keyl} are the sums of levels not lowered due to \hyperref[itm:catch]{\textbf{CTCH}} (\catchPercentage\ of candidates), \hyperref[itm:followingifs]{\textbf{IFS}} (\ifPercentage), and \hyperref[itm:withkeys]{\textbf{KEYL}} (\withKeysPercentage), respectively. Columns \textbf{cnds} and \textbf{keyr} are the sums of levels not transformed due to \hyperref[itm:wrapping]{\textbf{CNDS}} (\condPercentage) and not raised due to \hyperref[itm:withoutkeys]{\textbf{KEYR}} (\withoutKeysPercentage), respectively.

The discrepancy between \textbf{keyl} (\withKeysPercentage) and \textbf{keyr} (\withoutKeysPercentage) suggests that our tool is more frequently attempting to lower levels than raise them. This tendency may be due to few program elements changing at a specific time. As such, it is expected that levels would be lowered more often than they are raised, but \hyperref[itm:withkeys]{\textbf{KEYL}} is curbing the lowering. Thus, contrary to previous studies~\cite{Li2017}, log messages
may play a more significant role in determining log types than their placement. 

In summary, we filtered out 2,819 non-feature logs that are calculated by adding the total number of logs in columns \textbf{ctch}, \textbf{ifs}, \textbf{cnds}, \textbf{keyl}, \textbf{keyr}, and \textbf{inh}, and evaluated 1,154 feature logs, which account for 29.05\% of all candidate logging statements.
\paragraph{Transformation}

To ascertain transformation grade, we consider the distance between two adjacent levels to be 1. 
\takeaway{Transformations (avg.~\transDist\ levels) were not overly drastic yet far enough to be noticeable.}

To discover transformation directionality, columns \textbf{low} and \textbf{rse} depict the number of lowered (\lowPercentage)
and raised (\raisePercentage) log levels, respectively, stipulating that our tool more frequently lowers levels. As with columns \textbf{keyl} and \textbf{keyr}, this is most likely indicative of the relatively small number of features that developers focus on at a particular time.
\takeaway{Levels are typically lowered (\lowPercentage), potentially facilitating focus on fewer features of interest.}

\paragraph{Log Level Diversity}

Evident from column \textbf{$\sigma_{\mathit{pre}}$}, which is the stdev of log levels \emph{before} transformation,\footnote{Only includes JUL\@.
N/A indicates that JUL was not used.}---averaging only \distributionBefore---is that the full spectrum of available log levels is not always utilized. Column \textbf{$\sigma_{\mathit{post}}$}, on the other hand, is stdev \emph{after} transformation---averaging \distributionAfter.
\takeaway{Our tool increased logging level distribution by \distributionPercentageIncrease, thereby utilizing more level spectrum.}

\subsubsection{Qualitative Analysis}\label{sec:eval:qual} 




We discuss several instances where our tool did and did not work as intended. While evaluating \guava,
one manual modification to unit test code where a log level was being tested (i.e., \javainline{assertEquals(Level.INFO, record.getLevel())}) was necessary. Our tool transformed this tested log level to \ji{FINEST}, which failed the test suite. While more sophisticated analysis is needed to handle such cases, we conjecture they are rare; we only found one.

The following transformation occurred in \selenium. The color \textcolor{cadmiumred}{red} represents lines removed and \textcolor{emerald}{green} lines added:

\begin{diffcode}
if (!check.isAlive()) {
|\Vlabel{lne:infolog}|-	LOG.info("Node is not alive: " + check.getMessage());
|\Vlabel{lne:severelog}|+	LOG.severe("Node is not alive: " + check.getMessage());
 	// Throw an exception to force another check sooner.
|\Vlabel{lne:nodethrow}| 	throw new UnsupportedOperationException("Node can't..");}
\end{diffcode}

\noindent The logging statement at \lref{lne:infolog} indicates a failure before an exception throw on \lref{lne:nodethrow}. The level is erroneously \ji{INFO}. Because this code area is a ``hot spot,'' i.e., being either frequently or recently edited, fortunately, our tool fixed this level by transforming it to \ji{SEVERE} (\lref{lne:severelog}). This fix was later incorporated into \selenium's mainline~\cite{Anonymous2020}. 

%

The following transformation occurred in \CoreNLP: 

\begin{diffcode}
 if (o == null) {
|\Vlabel{lne:overview}| 	logger.severe("READER ERROR: Failed to find...");
|\Vlabel{lne:detail}|-	logger.severe("This happens because a few relation...");
|\Vlabel{lne:incorrect}|+	logger.fine("This happens because a few relation...");
\end{diffcode}

\noindent In place of using string concatenation in a single statement, \lref{lne:overview} logs an event overview, while \lref{lne:detail} logs the details. Although the heuristics worked for \lref{lne:overview}, they failed for \lref{lne:detail}, resulting in an incorrect transformation at \lref{lne:incorrect}.
This pattern was not observed in other subjects.

The following transformation took place in \jenkins:

\begin{diffcode}
|\label{lne:buildinfo}|-	LOGGER.log(INFO,"{0} main build action completed:{1}"..);
|\label{lne:buildfinest}|+	LOGGER.log(FINEST,"{0} main build action completed:{1}..");
\end{diffcode}

\noindent As the feature associated with the log at \lref{lne:buildinfo} was not of recent developer interest, our tool correctly lowered its level on \lref{lne:buildfinest}. This transformation, with developers expressing that, ``[it is p]robably a good idea: [i]t's time we started removing this from the general system log~\cite{Beck2019a},'' was also accepted into \jenkins' mainline. A subsequent comment for a similar transformation further motivated our approach by stating that, ``I [ha]ve [grew] so used to these messages over the years.'' 

\subsubsection{Bug Study}\label{sec:eval:bugs}

To answer~\ref{rq:bugs}, we mined software repositories for bug and non-bug fixes with nearby feature logs and applied our tool to versions leading up to the changesets.

\paragraph{Methodology}


To discover changesets having buggy and non-buggy logging contexts, we used \gitcproc~\cite{Casalnuovo2017}, a tool for processing and classifying Git commits that has been used previously~\cite{Gharbi2019,Tian2017,Khatchadourian2020} to automatically detect and analyze bug fixes in GitHub. Natural language processing (NLP) is used on commit log messages to identify bug fixes, and changesets can be queried for specific keywords. As we focus on lines with logging statements \emph{surrounding} modified lines, e.g., line~\ref{lne:finer1}, \cref{lst:wombat2}, we altered \gitcproc\ to search
\emph{outside} changesets, set the query keywords to those corresponding to logging APIs,\footnote{For succinctness, only JUL was used.}
and manually examined 
the results.


Once commits containing buggy and non-buggy feature logging statement contexts were identified, we ran our tool on the project version of the \emph{immediately preceding} commits. The tool examines up to 1000 commits in each project version's git history.  All heuristics in \cref{sec:heuristics} except for \hyperref[itm:threshold]{\textbf{TDIST}} were enabled to distinguish feature logging statements in contexts, \hyperref[itm:threshold]{\textbf{TDIST}} was set to \ji{INT_MAX}, and $\leq1\text{K}$ commits leading up to the preceding commits were processed. For each subject, $\leq\chosenCommitsFromEachContextType$ of each context type was chosen, such that the number of project versions considered was minimized.
This increased the likelihood of successful project building as the selected versions may have been intermediate.
Otherwise,
contexts were chosen randomly.
Only \bugsNumberOfSubjects\
subjects
were used as some had long running times by \gitcproc,
no logging contexts, or only unbuildable versions.


\begin{table}
    \scriptsize
    \centering
    \begin{tabular}{rlllll}
	\toprule
	&orig.~level&bug&\hyperref[itm:lower]{\textbf{LOW}}&\hyperref[itm:withoutkeys]{\textbf{KEYR}}&dir\\
	\midrule
	1&\jit{INFO}&\textit{T}&N/A&N/A&NONE\\
	2&\jit{FINEST}/\jit{FINER}/\jit{FINE}&\textit{T}&N/A&\textit{T}&RAISE\\
	3&\jit{FINEST}/\jit{FINER}/\jit{FINE}&\textit{T}&N/A&\textit{F}&NONE\\
	4&\jit{FINER}/\jit{FINE}/\jit{INFO}&\textit{F}&\textit{T}&N/A&NONE\\
	5&\jit{FINER}/\jit{FINE}/\jit{INFO}&\textit{F}&\textit{F}&N/A&LOWER\\
	6&\jit{FINEST}&\textit{F}&N/A&N/A&NONE\\
	\bottomrule
	\end{tabular}
	\caption{``Ideal'' level transformation directions for feature logs.
	Columns \textbf{orig.~level} and \textbf{bug} denote the original log level and whether the log is in a buggy context, respectively.
	Columns \textbf{\hyperref[itm:lower]{\textbf{LOW}}} and \textbf{\hyperref[itm:withoutkeys]{\textbf{KEYR}}} represent whether the statement passes any of 
	\hyperref[itm:catch]{\textbf{CTCH}}, \hyperref[itm:followingifs]{\textbf{IFS}}, \hyperref[itm:withkeys]{\textbf{KEYL}} and \hyperref[itm:withoutkeys]{\textbf{KEYR}}, respectively. These columns help to hone in the analysis on feature logs. 
	Finally, column \textbf{dir} portrays the ideal level direction. N/A is either \textit{T} or \textit{F}.
	}\label{tab:oracle}
\end{table}

To assess our tool's transformations when applied to project versions immediately preceding commits containing buggy and non-buggy feature logging contexts, we define an ``ideal'' level direction
to be used as an oracle (\cref{tab:oracle}).

For example,
\cref{lst:wombat2} fixes a bug accepting invalid temperatures. Here, \ji{FINER} logs at lines~\ref{lne:finer1}--\ref{lne:guardEnd} are in a \emph{buggy} context. Higher levels, e.g., \ji{FINE} (lines~\ref{lne:transformFine}--\ref{lne:transformFine2}, \cref{lst:wombat4}), may have helped bring attention to this bug \emph{earlier}, i.e., by documenting invalid temperature values \emph{more} prominently (row 2, \cref{tab:oracle}). Conversely, a \ji{FINE} log, e.g., line~\ref{lne:fileLog}, \cref{lst:wombat}, that is \emph{not} in a buggy context may have its level lowered (e.g., line~\ref{lne:fileDecrease}, \cref{lst:wombat4}) so that \emph{other} logs \emph{in} buggy contexts are more noticeable (row 5). Rows 1 and 6 are boundaries---logs in buggy contexts at the highest level cannot be raised and vice-versa. Logs in row 3 are in buggy contexts but do not have their levels raised due to
failing \hyperref[itm:withkeys]{\textbf{KEYL}}, while logs in row 4 are in non-buggy contexts but are not lowered due to
failing \hyperref[itm:lower]{\textbf{LOW}}.

\paragraph{Results}

\begin{table}[t]
	\centering
    \footnotesize{
	    \begin{tabular}{lllllllllll}
		\toprule
		subject&vers&\multicolumn{2}{c}{ctxts}&\multicolumn{2}{c}{rse}&\multicolumn{2}{c}{low}&\multicolumn{2}{c}{none}&i=a\\
		& & bug & $\neg$bug & idl & act & idl & act & idl & act & \\
		\cline{3-10}
		\midrule
		bc-java&8&0&10&0&0&2&2&8&8&10\\
		blueocean&6&7&3&4&2&0&0&6&8&6\\
		errorprone&2&0&2&0&0&0&0&2&2&2\\
		guava&11&1&14&0&0&1&0&14&15&14\\
		hollow&10&1&9&0&0&2&0&8&10&8\\
		IRCT&10&11&20&0&0&8&10&23&21&29\\
		JacpFX&4&3&1&0&0&0&1&4&3&3\\
		jenkins&5&9&22&4&0&3&2&24&29&24\\
		OpenGrok&6&28&20&0&1&6&12&42&35&35\\
		selenium&17&20&20&0&0&6&8&34&32&34\\
		SpotBugs&18&20&20&0&0&6&6&34&34&36\\
		WALA&4&1&3&1&0&0&0&3&4&3\\
		Total&101&101&144&9&3&34&41&202&201&204\\
		\cline{3-4}
		& & \multicolumn{2}{c}{245} & & & & & & & \\
		\bottomrule
	    \end{tabular}
	}
	\caption{Feature logging statements in change contexts.
	Column \textbf{vers} depicts the number of subject versions, column \textbf{ctxts} the buggy (\textbf{bug}) and non-buggy (\textbf{$\neg$bug}) logging contexts extracted, and columns \textbf{rse}, \textbf{low}, and \textbf{none} the feature log levels that are ideally (\textbf{idl}) and actual were (\textbf{act}) raised, lowered, and not altered, respectively. 
	Column \textbf{i=a} is the number of matching ideal and actual level directions.
	}\label{tab:bugres}
\end{table}

In \cref{tab:bugres}, 
column \textbf{ctxts} the buggy averages \num{1} per vers and non-buggy \textbf{$\neg$bug} averages \num{1.43} per vers. \takeaway{Levels of feature logs in change contexts were transformed in ideal directions \bugsIdealVersusActualRateApprox\ (\bugsIdealVersusActualFraction) of the time, potentially bringing problematic feature implementations into higher focus and exposing bugs.}

\paragraph{Discussion}

Higher/lower levels of feature logs in buggy/non-buggy contexts may have helped reveal/highlight problematic feature implementations, especially considering that such logs, e.g., lines~\ref{lne:finer1}--\ref{lne:guardEnd} of \cref{lst:wombat2}, typically include critical variables. Having these logs appear more prominently may induce fixes, e.g., lines~\ref{lne:guardStart} and~\ref{lne:except}.
Unnecessarily altering levels in equally essential to avoid false positives that introduce noise. As buggy code tends to be more frequently and recently edited~\cite{Mondal2017},
our approach is well-suited to ideally adjust---if necessary---feature logging statement levels in directions that may divulge bugs and avoid time-consuming fixes.

\subsubsection{Pull Request Study}\label{sec:eval:pulls}


To answer~\ref{rq:useful},
we submitted
pull requests containing our tool's transformations.

\paragraph{Results}

As of this writing, requests have been made to \pullRequestedProjects\ projects---
2 pull requests have been accepted, 5 have been refused, and 11 have been pending. Projects that merged transformations include \jenkins, a well-known continuous integration (CI) engine, having \num{\sim15}K stars and \num{\sim6.1}K forks on GitHub, and \selenium, a prominent web application testing and automation tool, having \num{\sim17}K stars and \num{\sim5.6}K forks.
As suggested by
these statistics, although only \pullRequestedProjectsAcceptedWord\ requests been accepted so far, the merged transformations have far-reaching impact as the projects include libraries and frameworks that are widely used in diverse circles---\selenium, for example, is used by \num{55931} other projects~\cite{GitHub2020a}.
Furthermore, the acceptance and rejection rates are comparable to that of previous work~\cite{Li2018a}.
\takeaway{\pullRequestedProjectsAccepted\ projects, both widely-used, having $\geq56\text{K}$ integrations---thus ensuring developer impact---accepted pull requests at rates comparable to previous work~\cite{Li2018a}.}

\paragraph{Discussion}

Our pull requests were rejected by five different projects of varying sizes. Developers from these projects, with the exception of the \guava\ developer, are more indifferent than developers from projects with accepted pull requests. We got a response that said, ``We are not interested in your tool. Please do not send any further PRs from this address.'' Furthermore, PRs with fewer changes are more likely to be accepted, whereas those with more changes are prone to be rejected.

Apparent during the study was that our approach encourages developers to actively consider how their logging statement levels evolve alongside their core software.
Feedback from rejected requests includes questions on whether or not our approach applies to very mature and largely stable projects that are in ``maintenance mode''~\cite{Baker2019a,Stewart2019}.
In this scenario,
developers respond to bug reports,
resulting in \emph{consistent} modifications to \emph{diverse} system modules. In such cases, our tool will never pick more ``interesting'' parts of the system as they are all equally or not equally ``interesting;'' application/system code that is under
active development may be more amenable.

A question~\cite{Baker2019} was also raised regarding the approach's applicability to library vs.~application code. Notably, parts of a library that are important to library developers may not match the interest level of application developers using the library. This problem---which may be more prevalent with public API implementations---touches on a broader issue of log composition, i.e., application developers' log intentions may not coincide with the application's dependencies in general.

\subsection{Threats to Validity}

Subjects may not be representative of real-world
log usage. For mitigation, subjects were chosen from diverse domains and sizes, as well as those used in previous studies~\cite{Khatchadourian2019,Ketkar2019}.

Git repositories with very large commits may be too coarse-grained to detect program element modifications accurately.
This limitation is standard among approaches that mine
software repositories. Furthermore, our keyword related heuristics cannot work with typos.
There were still \logsNotAlteredDueToKeywords\ logging statements in our study whose levels were not altered due to keywords, suggesting that typos are not pervasive. Adding support for the use of misspelled keywords is straight-forward.

Moreover, during the pull request study (\cref{sec:eval:pulls}), there was at least one instance where developers desired to modify logging statement levels eventually but had not yet done so.

We may miss some of these ``wrapped'' logs that are excluded by the \textbf{CNDS} heuristic. However, we did not find a significant number of such logs during our evaluation (only ${\sim}\SI{6.3}{\percent}$). Moreover, such logs are likely to reside in developer-defined library code that is used by many features, making it difficult to correlate DOI.\@ Lastly, we understand that the pattern involving \textbf{CNDS} is used to improve performance. However, with the relatively recent introduction of logging API accepting $\lambda$-expression, we expect that developers will increasingly use those instead of the pattern.

\section{Related Work}

Logging frameworks may include logger hierarchies~\cite{Oracle2018a,ASF2020}, where log objects can be designated to log certain features implementations.
Individual loggers can thus be enabled or disabled to facilitate focus on particular feature implementations. However, because they are typically many features
\begin{enumerate*}[(i)]
    \item whose interests change over time~\cite{Kersten2006} and
    \item whose implementations are not localized~\cite{Kiczales1997},
\end{enumerate*}
developers are still burdened with manually maintaining logger hierarchies.







\citet{Li2017} determine log levels for \emph{new} logging statements to be consistent with other logging practices within the same project. On the other hand, we focus on log level \emph{evolution} and how levels relate to feature interest. While their work can be retargeted for evolution by treating the logging statement under consideration as ``new'' and subsequently predicting a possibly new level, the goals of their approach are quite different from ours. Firstly, they focus on \emph{general} logging statements, whereas we focus on feature logs.
Secondly, they aim to predict a level that fits well with the current \emph{project} trends, including log placement, existing logs in the same file, and message content. On the contrary, our goal is to better align feature logging statement levels with current \emph{feature} interest, which may have little bearing on placement.
In other words, feature interest at a particular point in time may be well-independent of the logging practices previously employed.

\citet{Chen2017,Hassani2018} detect and correct mismatches between log messages and levels. However, they do not consider how varying developer interests in particular features affect logging levels over time. While these approaches are useful in discovering error logging statements that use lower-than-normal logging levels, they may not be as useful for event-type logs, which may be more tied to features.

\citet{Li2018a} predict log revisions by mining the correlation between logging context and code modifications with the premise that logging code in similar contexts deserves similar modifications. As far as we can tell, however, they do not explicitly deal with logging levels. \citet{Li2017a,Kabinna2018} determine the likelihood of log change
but do not suggest a specific modification. \citet{Li2018} predict---using topic modeling---the likelihood that a particular code snippet should including logging. \citet{Shang2014} examine log lines using development knowledge to resolve log inquiries, \citet{Yuan2012b} add appropriate logging statements to enhance failure diagnosis, and \citet{Zhu2019} evaluate log parsers.

Several approaches combat information overload. \citet{Haas2020} use static analysis to detect unnecessary source code. \citet{Fowkes2017} introduce auto-folding for source code summarization. Information overload is also an issue for logging, and many approaches enhance logging statements by, e.g., determining log placement~\cite{Fu2014,Zhu2015}, optimizing the number of logging statements~\cite{Lal2017}, and enriching log messages~\cite{Yuan2012a,He2018,Li2019}.
\citet{Xu2009} mine log output to detect problems.



\insertrefRKi{\cite{Casalnuovo2017,Casalnuovo2015}}

\citet{Khatchadourian2017a} also integrate the Mylyn DOI model but for Aspect-Oriented Programming (AOP). To the best of our knowledge, we are the first to manipulate a DOI model using software repository mining programmatically. Other work mines software repositories for evolution~\cite{Kagdi2007}, detecting refactorings~\cite{Tsantalis2018}, and design flaw detection~\cite{Ratiu2004}. Additionally, approaches support software evolution more generally, e.g., by refactoring programs to use enumerated types~\cite{Khatchadourian2017b}, default methods~\cite{Khatchadourian2017}, and lambda expressions~\cite{Ketkar2019}.
\citet{Bhatta2012} also use graphs for software evolution.


\section{Conclusion \& Future Work}\label{sec:conc}

Our automated approach ``reinvigorates'' feature logging statement levels based on ``interest'' of surrounding code as determined by software repository mining. Distinguishing feature logs is performed via introduced heuristics. The approach is implemented as an Eclipse IDE plug-in, using JGit and Mylyn, and evaluated on \NumProjects\ projects with \MLOC\ MLOC and \logsThousands\ logging statements. Our tool successfully analyzes \analyzedPercentage\ of logging statements, increases log level distributions by \distributionPercentageIncreaseApprox, 
increases the focus of logs within bug fix contexts \bugsIdealVersusActualRateApprox\ of the time, and integrated transformations into several
projects. In the future, we will explore leveraging existing Mylyn task contexts, expanding the heuristics,
and surveying developers. In addition, we will enhance our algorithm to partition DOI ranges by using Machine Learning approaches.

\section*{Acknowledgments}


We would particularly like to thank all the developers that offered extremely helpful feedback through our pull requests. These include Manu Sridharan, Daniel Beck, and David P.~Baker. We would also like to thank Oren Friedman and Walee Ahmed for their help with experiments and infrastructure, respectively, as well as Shigeru Chiba, Hidehiko Masuhara, and Iulian Neamtiu for their thoughtful insight and feedback. Support for this project was provided by PSC-CUNY Awards \#617930049 and \#638010051, jointly funded by The Professional Staff Congress and The City University of New York.

\printbibliography%

@String{aosd     = {International Conference on Aspect-Oriented Software Development}}

@String{apache   = {Apache}}

@String{ase      = {International Conference on Automated Software Engineering}}

@String{ecoop    = {European Conference on Object-Oriented Programming}}

@String{ese      = {Empirical Softw. Engg.}}

@String{fase     = {International Conference on Fundamental Approaches to Software Engineering}}

@String{fse      = {ACM Symposium on the Foundations of Software Engineering}}

@String{icse     = {International Conference on Software Engineering}}

@String{icsm     = {International Conference on Software Maintenance}}

@String{icsme    = {International Conference on Software Maintenance and Evolution}}

@String{issta    = {International Symposium on Software Testing and Analysis}}

@String{oopsla   = {ACM SIGPLAN International Conference on Object-Oriented Programming, Systems, Languages, and Applications}}

@String{springer = {Springer-Verlag}}

@String{toplas   = {ACM Transactions on Programming Languages and Systems}}

@String{tse      = {IEEE Transactions on Software Engineering}}

@InProceedings{Kersten2006,
  author    = {Kersten, Mik and Murphy, Gail C.},
  title     = {Using Task Context to Improve Programmer Productivity},
  booktitle = fse,
  year      = {2006},
  series    = {SIGSOFT '06/FSE-14},
  publisher = {ACM},
  location  = {Portland, Oregon, USA},
  isbn      = {1-59593-468-5},
  pages     = {1--11},
  doi       = {10.1145/1181775.1181777},
  acmid     = {1181777},
  address   = {New York, NY, USA},
  numpages  = {11},
}

@Article{Li2017,
  author     = {Li, Heng and Shang, Weiyi and Hassan, Ahmed E.},
  title      = {Which Log Level Should Developers Choose for a New Logging Statement?},
  journal    = ese,
  year       = {2017},
  volume     = {22},
  number     = {4},
  month      = aug,
  pages      = {1684--1716},
  issn       = {1382-3256},
  doi        = {10.1007/s10664-016-9456-2},
  address    = {USA},
  issue_date = {August 2017},
  numpages   = {33},
  publisher  = {Kluwer Academic Publishers},
}

@Online{Oracle2018a,
  author    = {Oracle},
  title     = {{Logger (Java SE 10 \& JDK 10)}},
  year      = {2018},
  url       = {http://docs.oracle.com/javase/10/docs/api/java/util/logging/Logger.html},
  urldate   = {2020-02-29},
}

@Online{OracleCatch,
  author    = {Oracle},
  title     = {The catch Blocks},
  year      = {2021},
  url       = {https://docs.oracle.com/javase/tutorial/essential/exceptions/catch.html},
  urldate   = {2021-06-15},
}

@InProceedings{Khatchadourian2017,
  author    = {Khatchadourian, Raffi and Masuhara, Hidehiko},
  title     = {Automated Refactoring of Legacy Java Software to Default Methods},
  booktitle = icse,
  year      = {2017},
  series    = {ICSE '17},
  publisher = {IEEE Press},
  location  = {Buenos Aires, Argentina},
  isbn      = {978-1-5386-3868-2},
  pages     = {82--93},
  doi       = {10.1109/ICSE.2017.16},
  acmid     = {3097379},
  address   = {Piscataway, NJ, USA},
  numpages  = {12},
}

@Article{Li2018,
  author    = {Li, Heng and Chen, Tse-Hsun and Shang, Weiyi and Hassan, Ahmed E.},
  title     = {Studying software logging using topic models},
  journal   = {Empirical Software Engineering},
  year      = {2018},
  volume    = {23},
  number    = {5},
  month     = oct,
  pages     = {2655--2694},
  issn      = {1573-7616},
  doi       = {10.1007/s10664-018-9595-8},
  day       = {01},
}

@Article{Yuan2012a,
  author     = {Yuan, Ding and Zheng, Jing and Park, Soyeon and Zhou, Yuanyuan and Savage, Stefan},
  title      = {Improving Software Diagnosability via Log Enhancement},
  journal    = {ACM Trans. Comput. Syst.},
  year       = {2012},
  volume     = {30},
  number     = {1},
  month      = feb,
  pages      = {4:1--4:28},
  issn       = {0734-2071},
  doi        = {10.1145/2110356.2110360},
  acmid      = {2110360},
  address    = {New York, NY, USA},
  articleno  = {4},
  issue_date = {February 2012},
  numpages   = {28},
  publisher  = {ACM},
}

@InProceedings{Li2018a,
  author    = {Li, Shanshan and Niu, Xu and Jia, Zhouyang and Wang, Ji and He, Haochen and Wang, Teng},
  title     = {Logtracker: Learning Log Revision Behaviors Proactively from Software Evolution History},
  booktitle = {International Conference on Program Comprehension},
  year      = {2018},
  series    = {ICPC '18},
  publisher = {ACM},
  location  = {Gothenburg, Sweden},
  isbn      = {978-1-4503-5714-2},
  pages     = {178--188},
  doi       = {10.1145/3196321.3196328},
  acmid     = {3196328},
  address   = {New York, NY, USA},
  numpages  = {11},
}

@InProceedings{Bhatta2012,
  author    = {Bhattacharya, Pamela and Iliofotou, Marios and Neamtiu, Iulian and Faloutsos, Michalis},
  title     = {Graph-based Analysis and Prediction for Software Evolution},
  booktitle = icse,
  year      = {2012},
  series    = {ICSE '12},
  publisher = {IEEE Press},
  location  = {Zurich, Switzerland},
  isbn      = {978-1-4673-1067-3},
  pages     = {419--429},
  url       = {http://dl.acm.org/citation.cfm?id=2337223.2337273},
  acmid     = {2337273},
  address   = {Piscataway, NJ, USA},
  numpages  = {11},
}

@InProceedings{Xu2009,
  author       = {Xu, Wei and Huang, Ling and Fox, Armando and Patterson, David and Jordan, Michael I.},
  title        = {Detecting Large-scale System Problems by Mining Console Logs},
  booktitle    = {Symposium on Operating Systems Principles},
  year         = {2009},
  series       = {SOSP '09},
  organization = {SIGOPS},
  publisher    = {ACM},
  location     = {Big Sky, Montana, USA},
  isbn         = {978-1-60558-752-3},
  pages        = {117--132},
  doi          = {10.1145/1629575.1629587},
  acmid        = {1629587},
  address      = {New York, NY, USA},
  numpages     = {16},
}

@InProceedings{Syer2013,
  author    = {M. D. Syer and Z. M. Jiang and M. Nagappan and A. E. Hassan and M. Nasser and P. Flora},
  title     = {Leveraging Performance Counters and Execution Logs to Diagnose Memory-Related Performance Issues},
  booktitle = icsm,
  year      = {2013},
  month     = sep,
  pages     = {110--119},
  doi       = {10.1109/ICSM.2013.22},
  issn      = {1063-6773},
}

@InProceedings{Zhu2015,
  author    = {Zhu, Jieming and He, Pinjia and Fu, Qiang and Zhang, Hongyu and Lyu, Michael R. and Zhang, Dongmei},
  title     = {Learning to Log: Helping Developers Make Informed Logging Decisions},
  booktitle = icse,
  year      = {2015},
  series    = {ICSE '15},
  publisher = {IEEE Press},
  location  = {Florence, Italy},
  isbn      = {978-1-4799-1934-5},
  pages     = {415--425},
  url       = {http://dl.acm.org/citation.cfm?id=2818754.2818807},
  acmid     = {2818807},
  address   = {Piscataway, NJ, USA},
  numpages  = {11},
}

@InProceedings{Shang2014,
  author    = {W. {Shang} and M. {Nagappan} and A. E. {Hassan} and Z. M. {Jiang}},
  title     = {Understanding Log Lines Using Development Knowledge},
  booktitle = icsme,
  year      = {2014},
  month     = sep,
  pages     = {21--30},
  doi       = {10.1109/ICSME.2014.24},
  issn      = {1063-6773},
}

@Article{Zeng2019,
  author    = {Zeng, Yi and Chen, Jinfu and Shang, Weiyi and Chen, Tse-Hsun},
  title     = {Studying the characteristics of logging practices in mobile apps: a case study on F-Droid},
  journal   = ese,
  year      = {2019},
  number    = {24},
  pages     = {3394--3438},
  month     = feb,
  issn      = {1573-7616},
  day       = {07},
  doi       = {10.1007/s10664-019-09687-9},
}

@InProceedings{Lal2017,
  author    = {Sangeeta Lal and Neetu Sardana and Ashish Sureka},
  title     = {Analysis and Prediction of Log Statement in Open Source {Java} Projects},
  booktitle = {Doctoral Consortium at the International Conference on Open Source Systems},
  year      = {2017},
  location  = {Buenos Aires, Argentina},
  month     = may,
  pages     = {65--80},
}

@InProceedings{Fu2014,
  author    = {Fu, Qiang and Zhu, Jieming and Hu, Wenlu and Lou, Jian-Guang and Ding, Rui and Lin, Qingwei and Zhang, Dongmei and Xie, Tao},
  title     = {Where Do Developers Log? An Empirical Study on Logging Practices in Industry},
  booktitle = {Companion Proceedings of the 36\textsuperscript{th} International Conference on Software Engineering},
  year      = {2014},
  series    = {ICSE Companion 2014},
  publisher = {ACM},
  location  = {Hyderabad, India},
  isbn      = {978-1-4503-2768-8},
  pages     = {24--33},
  doi       = {10.1145/2591062.2591175},
  acmid     = {2591175},
  address   = {New York, NY, USA},
  numpages  = {10},
}

@Article{Hassani2018,
  author    = {Hassani, Mehran and Shang, Weiyi and Shihab, Emad and Tsantalis, Nikolaos},
  title     = {Studying and detecting log-related issues},
  journal   = {Empirical Software Engineering},
  year      = {2018},
  month     = mar,
  issn      = {1573-7616},
  doi       = {10.1007/s10664-018-9603-z},
  day       = {15},
}

@Article{Fowkes2017,
  author     = {Fowkes, Jaroslav and Chanthirasegaran, Pankajan and Ranca, Razvan and Allamanis, Miltiadis and Lapata, Mirella and Sutton, Charles},
  title      = {Autofolding for Source Code Summarization},
  journal    = tse,
  year       = {2017},
  volume     = {43},
  number     = {12},
  pages      = {1095--1109},
  month      = dec,
  issn       = {0098-5589},
  doi        = {10.1109/TSE.2017.2664836},
  issue_date = {December 2017},
  numpages   = {15},
  publisher  = {IEEE Press},
}

@article{Haas2020,
  author = {Haas, Roman and Niedermayr, Rainer and Roehm, Tobias and Apel, Sven},
  title = {Is Static Analysis Able to Identify Unnecessary Source Code?},
  year = {2020},
  issue_date = {February 2020},
  publisher = {Association for Computing Machinery},
  address = {New York, NY, USA},
  volume = {29},
  number = {1},
  issn = {1049-331X},
  doi = {10.1145/3368267},
  journal = {ACM Trans. Softw. Eng. Methodol.},
  month = jan,
  articleno = {Article 6},
  numpages = {23}
}

@Inproceedings{Rozinat2005,
  author    = {Rozinat, A. and van der Aalst, W. M. P.},
  title     = {Conformance Testing: Measuring the Fit and Appropriateness of Event Logs and Process Models},
  booktitle = {International Conference on Business Process Management},
  year      = {2005},
  series    = {BPM '05},
  pages     = {163--176},
  address   = {Berlin, Heidelberg},
  publisher = {Springer-Verlag},
  doi       = {10.1007/11678564_15},
  isbn      = {3540325956},
  location  = {Nancy, France},
  numpages  = {14},
}

@Article{Kabinna2018,
  author     = {Kabinna, Suhas and Bezemer, Cor-Paul and Shang, Weiyi and Syer, Mark D. and Hassan, Ahmed E.},
  title      = {Examining the Stability of Logging Statements},
  journal    = ese,
  year       = {2018},
  volume     = {23},
  number     = {1},
  pages      = {290--333},
  month      = feb,
  issn       = {1382-3256},
  address    = {USA},
  doi        = {10.1007/s10664-017-9518-0},
  issue_date = {February 2018},
  numpages   = {44},
  publisher  = {Kluwer Academic Publishers},
}

@Inproceedings{Tan2008,
  author    = {Tan, Jiaqi and Pan, Xinghao and Kavulya, Soila and Gandhi, Rajeev and Narasimhan, Priya},
  title     = {SALSA: Analyzing Logs as State Machines},
  booktitle = {USENIX Conference on Analysis of System Logs},
  year      = {2008},
  series    = {WASL '08},
  pages     = {6},
  address   = {USA},
  publisher = {USENIX Association},
  location  = {San Diego, California},
  numpages  = {1},
}

@InProceedings{Chen2017,
  author    = {Chen, Boyuan and Jiang, Zhen Ming},
  title     = {Characterizing and Detecting Anti-Patterns in the Logging Code},
  booktitle = icse,
  year      = {2017},
  series    = {ICSE '17},
  publisher = {IEEE Press},
  location  = {Buenos Aires, Argentina},
  isbn      = {9781538638682},
  pages     = {71--81},
  doi       = {10.1109/ICSE.2017.15},
  numpages  = {11},
}

@InProceedings{Kersten2005,
  author        = {Kersten, Mik and Murphy, Gail C.},
  title         = {{Mylar: a degree-of-interest model for IDEs}},
  booktitle     = aosd,
  year          = {2005},
  publisher     = {ACM},
  location      = {Chicago, Illinois},
  isbn          = {1-59593-042-6},
  pages         = {159--168},
  doi           = {10.1145/1052898.1052912},
  acmid         = {1052912},
  address       = {New York, NY, USA},
  numpages      = {10},
}

@WWW{EclipseFoundation2020a,
  author    = {{Eclipse Foundation, Inc.}},
  title     = {{Eclipse} {IDE}s},
  year      = {2020},
  url       = {http://eclip.se/gD},
  month     = aug,
  urldate   = {2020-03-02},
}

@Online{Oracle2018b,
  author    = {Oracle},
  title     = {{java.util.logging (Java SE 10 \& JDK 10)}},
  year      = {2018},
  url       = {http://docs.oracle.com/javase/10/docs/api/java/util/logging/package-summary.html},
  urldate   = {2020-03-02},
}

@Online{QOS.ch2019,
  author    = {QOS.ch},
  title     = {{SLF4J}},
  year      = {2019},
  url       = {http://www.slf4j.org},
  subtitle  = {{Simple Logging Facade for Java}},
  urldate   = {2020-03-02},
}

@Online{EclipseFoundation2020b,
  author    = {{Eclipse Foundation, Inc.}},
  title     = {{JGit}},
  year      = {2020},
  url       = {http://eclip.se/gF},
  urldate   = {2020-03-02},
}

@InProceedings{He2018,
  author    = {He, Pinjia and Chen, Zhuangbin and He, Shilin and Lyu, Michael R.},
  title     = {Characterizing the Natural Language Descriptions in Software Logging Statements},
  booktitle = ase,
  year      = {2018},
  series    = {ASE '18},
  publisher = {ACM},
  location  = {Montpellier, France},
  isbn      = {9781450359375},
  pages     = {178--189},
  doi       = {10.1145/3238147.3238193},
  address   = {New York, NY, USA},
  numpages  = {12},
}

@Online{QOS.ch2019a,
  author    = {QOS.ch},
  title     = {SLF4J Manual},
  year      = {2019},
  url       = {http://www.slf4j.org/manual.html#typical_usage},
  subtitle  = {Typical usage pattern},
  urldate   = {2020-03-03},
}

@InProceedings{Zhu2019,
  author    = {Zhu, Jieming and He, Shilin and Liu, Jinyang and He, Pinjia and Xie, Qi and Zheng, Zibin and Lyu, Michael R.},
  title     = {Tools and Benchmarks for Automated Log Parsing},
  booktitle = {International Conference on Software Engineering: Software Engineering in Practice},
  year      = {2019},
  series    = {ICSE-SEIP '19},
  publisher = {IEEE Press},
  location  = {Montreal, Quebec, Canada},
  pages     = {121--130},
  doi       = {10.1109/ICSE-SEIP.2019.00021},
  numpages  = {10},
}

@InProceedings{Li2019,
  author    = {Li, Zhenhao and Chen, Tse-Hsun and Yang, Jinqiu and Shang, Weiyi},
  title     = {{DLFinder}: Characterizing and Detecting Duplicate Logging Code Smells},
  booktitle = icse,
  year      = {2019},
  series    = {ICSE '19},
  publisher = {IEEE Press},
  location  = {Montreal, Quebec, Canada},
  pages     = {152--163},
  doi       = {10.1109/ICSE.2019.00032},
  numpages  = {12},
}

@InProceedings{Ratiu2004,
  author    = {Ratiu, Daniel and Ducasse, St\'{e}phane and Gundefinedrba, Tudor and Marinescu, Radu},
  title     = {Using History Information to Improve Design Flaws Detection},
  booktitle = {European Conference on Software Maintenance and Reengineering},
  year      = {2004},
  series    = {CSMR '04},
  publisher = {IEEE},
  isbn      = {076952107X},
  pages     = {223--232},
  doi       = {10.1109/csmr.2004.1281423},
  numpages  = {10},
}

@InProceedings{Yuan2012b,
  author       = {Ding Yuan and Soyeon Park and Peng Huang and Yang Liu and Michael M. Lee and Xiaoming Tang and Yuanyuan Zhou and Stefan Savage},
  title        = {Be Conservative: Enhancing Failure Diagnosis with Proactive Logging},
  booktitle    = {Operating Systems Design and Implementation},
  year         = {2012},
  series       = {OSDI '12},
  organization = {USENIX},
  publisher    = {USENIX Association},
  location     = {Hollywood, CA, USA},
  isbn         = {9781931971966},
  pages        = {293--306},
  address      = {USA},
  numpages     = {14},
}

@Article{Khatchadourian2017a,
  author     = {Khatchadourian, Raffi and Rashid, Awais and Masuhara, Hidehiko and Watanabe, Takuya},
  title      = {Detecting Broken Pointcuts Using Structural Commonality and Degree of Interest},
  journal    = {Sci. Comput. Program.},
  year       = {2017},
  volume     = {150},
  number     = {C},
  month      = dec,
  pages      = {56--74},
  issn       = {0167-6423},
  doi        = {10.1016/j.scico.2017.06.011},
  address    = {USA},
  issue_date = {December 2017},
  numpages   = {19},
  publisher  = {Elsevier North-Holland, Inc.},
}

@Article{Khatchadourian2017b,
  author    = {Raffi Khatchadourian},
  title     = {Automated refactoring of legacy {Java} software to enumerated types},
  journal   = {Automated Software Engineering},
  year      = {2017},
  volume    = {24},
  number    = {4},
  month     = dec,
  pages     = {757--787},
  issn      = {0928-8910},
  doi       = {10.1007/s10515-016-0208-8},
  day       = {01},
  publisher = {Springer US},
}

@InProceedings{Khatchadourian2019,
  author       = {Khatchadourian, Raffi and Tang, Yiming and Bagherzadeh, Mehdi and Ahmed, Syed},
  title        = {Safe Automated Refactoring for Intelligent Parallelization of {Java} 8 Streams},
  booktitle    = icse,
  year         = {2019},
  series       = {ICSE '19},
  organization = {ACM/IEEE},
  publisher    = {IEEE Press},
  location     = {Montr\'eal, QC, Canada},
  month        = may,
  pages        = {619--630},
  doi          = {10.1109/ICSE.2019.00072},
  acmid        = {3339586},
  address      = {Piscataway, NJ, USA},
  numpages     = {12},
  slides       = {http://www.slideshare.net/khatchad/safe-automated-refactoring-for-intelligent-parallelization-of-java-8-streams},
  tool         = {http://github.com/ponder-lab/Optimize-Java-8-Streams-Refactoring},
}

@Online{McDonald2018,
  author       = {Josh McDonald},
  title        = {{blueocean-plugin/DownstreamJobListener.java at 522fd48 · jenkinsci/blueocean-plugin}},
  year         = {2018},
  date         = {2018-01-24},
  url          = {http://git.io/JvVQ0},
  organization = {Jenkins},
  urldate      = {2020-03-05},
}

@Online{Decker2014,
  author       = {Colin Decker},
  title        = {{guava/EventBus.java at 0cd4e9f · google/guava}},
  year         = {2014},
  date         = {2014-08-05},
  url          = {http://git.io/JvV7P},
  organization = {Google},
  urldate      = {2020-03-05},
}

@Unpublished{Baeumer2001,
  author    = {B{\"a}umer, Dirk and Gamma, Erich and Kiezun, Adam},
  title     = {Integrating refactoring support into a {J}ava development tool},
  year      = {2001},
  date      = {2001},
  month     = oct,
  url       = {http://people.csail.mit.edu/akiezun/companion.pdf},
  booktitle = oopsla,
}

@InProceedings{Ketkar2019,
  author    = {Ketkar, Ameya and Mesbah, Ali and Mazinanian, Davood and Dig, Danny and Aftandilian, Edward},
  title     = {Type Migration in Ultra-large-scale Codebases},
  booktitle = icse,
  year      = {2019},
  series    = {ICSE '19},
  publisher = {IEEE Press},
  location  = {Montreal, Quebec, Canada},
  pages     = {1142--1153},
  doi       = {10.1109/ICSE.2019.00117},
  acmid     = {3339648},
  address   = {Piscataway, NJ, USA},
  numpages  = {12},
}

@Article{Li2017a,
  author     = {Heng Li and Weiyi Shang and Ying Zou and Ahmed Hassan},
  title      = {Towards Just-in-Time Suggestions for Log Changes},
  journal    = {Empirical Softw. Engg.},
  year       = {2017},
  volume     = {22},
  number     = {4},
  month      = aug,
  pages      = {1831--1865},
  issn       = {1382-3256},
  doi        = {10.1007/s10664-016-9467-z},
  address    = {USA},
  issue_date = {August 2017},
  numpages   = {35},
  publisher  = {Kluwer Academic Publishers},
}

@Article{Kagdi2007,
  author     = {Kagdi, Huzefa and Collard, Michael L. and Maletic, Jonathan I.},
  title      = {A Survey and Taxonomy of Approaches for Mining Software Repositories in the Context of Software Evolution},
  journal    = {J. Softw. Maint. Evol.},
  year       = {2007},
  volume     = {19},
  number     = {2},
  month      = mar,
  pages      = {77--131},
  issn       = {1532-060X},
  doi        = {10.1002/smr.344},
  address    = {USA},
  issue_date = {March 2007},
  numpages   = {55},
  publisher  = {John Wiley \& Sons, Inc.},
}

@InProceedings{Tsantalis2018,
  author    = {Tsantalis, Nikolaos and Mansouri, Matin and Eshkevari, Laleh M. and Mazinanian, Davood and Dig, Danny},
  title     = {Accurate and Efficient Refactoring Detection in Commit History},
  booktitle = icse,
  year      = {2018},
  series    = {ICSE '18},
  publisher = {ACM},
  location  = {Gothenburg, Sweden},
  isbn      = {978-1-4503-5638-1},
  pages     = {483--494},
  doi       = {10.1145/3180155.3180206},
  acmid     = {3180206},
  address   = {New York, NY, USA},
  numpages  = {12},
}

@Software{Tang2020,
  author    = {Anonymous},
  title     = {{Reinvigorate Logging Levels}},
  year      = {2020},
  version   = {v1.9.0},
  note      = {Anonymized repository},
  url       = {https://anonymous.4open.science/r/d301551f-765a-4825-9f01-79d570e0cc94},
  urldate   = {2020-06-17},
}

@InProceedings{Casalnuovo2017,
  author    = {Casalnuovo, Casey and Suchak, Yagnik and Ray, Baishakhi and Rubio-Gonz\'{a}lez, Cindy},
  title     = {GitcProc: A Tool for Processing and Classifying GitHub Commits},
  booktitle = issta,
  year      = {2017},
  series    = {ISSTA '17},
  publisher = {ACM},
  location  = {Santa Barbara, CA, USA},
  isbn      = {978-1-4503-5076-1},
  pages     = {396--399},
  doi       = {10.1145/3092703.3098230},
  acmid     = {3098230},
  address   = {New York, NY, USA},
  numpages  = {4},
}

@InProceedings{Casalnuovo2015,
  author    = {Casalnuovo, Casey and Devanbu, Prem and Oliveira, Abilio and Filkov, Vladimir and Ray, Baishakhi},
  title     = {Assert Use in GitHub Projects},
  booktitle = icse,
  year      = {2015},
  series    = {ICSE '15},
  publisher = {IEEE Press},
  location  = {Florence, Italy},
  isbn      = {978-1-4799-1934-5},
  pages     = {755--766},
  url       = {http://dl.acm.org/citation.cfm?id=2818754.2818846},
  acmid     = {2818846},
  address   = {Piscataway, NJ, USA},
  numpages  = {12},
}

@WWW{Davis2014,
  author       = {Sam Davis},
  title        = {433030 -- Add to Task Context does not work},
  year         = {2014},
  date         = {2014-04-17},
  url          = {http://eclip.se/gI},
  organization = {Eclipse Foundation},
  urldate      = {2020-06-08},
}

@WWW{Pilgrim2014,
  author       = {Jens von Pilgrim},
  title        = {Create context from git diff},
  year         = {2014},
  date         = {2014-04-17},
  url          = {http://eclip.se/gJ},
  subtitle     = {Mylyn},
  titleaddon   = {Eclipse Community Forums},
  organization = {Eclipse Foundation},
  urldate      = {2020-06-08},
}

@WWW{GitHub2020,
  author     = {{GitHub, Inc.}},
  title      = {Building Apps},
  year       = {2020},
  url        = {http://git.io/JfS9E},
  titleaddon = {GitHub Developer Guide},
  urldate    = {2020-06-08},
}

@InProceedings{Alizadeh2019,
  author    = {Alizadeh, Vahid and Ouali, Mohamed Amine and Kessentini, Marouane and Chater, Meriem},
  title     = {RefBot: Intelligent Software Refactoring Bot},
  booktitle = ase,
  year      = {2019},
  series    = {ASE '19},
  publisher = {IEEE Press},
  location  = {San Diego, California},
  isbn      = {9781728125084},
  pages     = {823--834},
  doi       = {10.1109/ASE.2019.00081},
  numpages  = {12},
}

@WWW{GitHub2020a,
  author    = {{GitHub, Inc.}},
  title     = {Network Dependents~\textbullet~SeleniumHQ/selenium},
  year      = {2020},
  url       = {http://git.io/JfS9O},
  urldate   = {2020-06-10},
}

@WWW{EclipseFoundation2020,
  author    = {{Eclipse Foundation, Inc.}},
  title     = {Eclipse {Mylyn} Open-Source Project},
  year      = {2020},
  url       = {http://eclip.se/gC},
  urldate   = {2020-06-12},
}

@InProceedings{Kiczales1997,
  author    = {Kiczales, Gregor and Lamping, John and Mendhekar, Anurag and Maeda, Chris and Lopes, Cristina and Loingtier, Jean-Marc and Irwin, John},
  title     = {Aspect-oriented programming},
  booktitle = ecoop,
  year      = {1997},
  publisher = {Springer},
  isbn      = {978-3-540-69127-3},
  pages     = {220--242},
  doi       = {10.1007/bfb0053381},
}

@WWW{ASF2020,
  author     = {{Apache Software Foundation}},
  title      = {Log4j},
  year       = {2020},
  url        = {http://logging.apache.org/log4j/2.x/manual/architecture.html#Logger_Hierarchy},
  titleaddon = {Log4j 2 Architecture},
  urldate    = {2020-06-12},
}

@InProceedings{Chang2008,
  author    = {Chang, Hung-Fu and Mockus, Audris},
  title     = {Evaluation of Source Code Copy Detection Methods on {FreeBSD}},
  booktitle = {International Working Conference on Mining Software Repositories},
  year      = {2008},
  series    = {MSR '08},
  publisher = {ACM},
  location  = {Leipzig, Germany},
  isbn      = {9781605580241},
  pages     = {61--66},
  doi       = {10.1145/1370750.1370766},
  address   = {New York, NY, USA},
  numpages  = {6},
}

@WWW{Anonymous2020,
  author       = {Anonymous},
  title        = {Rejuvenate log levels by anonymous~\textbullet~Pull Request \#7737~\textbullet~SeleniumHQ/selenium},
  year         = {2020},
  date         = {2020-02-18},
  url          = {http://git.io/Jfd2W},
  note         = {Following this URL may reveal author identity},
  organization = {Selenium},
  urldate      = {2020-06-17},
}

@WWW{Baker2019,
  author       = {David P. Baker},
  title        = {Rejuvenate log levels by anonymous~\textbullet~Pull Request \#3713~\textbullet~google/guava},
  year         = {2019},
  date         = {2019-11-20},
  url          = {http://git.io/Jfdae},
  note         = {Following this URL may reveal author identity},
  organization = {Google},
  urldate      = {2020-07-27},
}

@WWW{Baker2019a,
  author       = {David P. Baker},
  title        = {Rejuvenate logging statement levels by anonymous~\textbullet~Pull Request \#3435~\textbullet~google/guava},
  year         = {2019},
  date         = {2019-04-05},
  url          = {http://git.io/Jfdal},
  note         = {Following this URL may reveal author identity},
  organization = {Google},
  urldate      = {2020-06-17},
}

@WWW{Stewart2019,
  author       = {Simon Stewart},
  title        = {Rejuvenate log levels by anonymous~\textbullet~Pull Request \#7170~\textbullet~SeleniumHQ/selenium},
  year         = {2019},
  date         = {2019-05-07},
  url          = {http://git.io/JfdaA},
  note         = {Following this URL may reveal author identity},
  organization = {Selenium},
  urldate      = {2020-06-17},
}

@WWW{Beck2019a,
  author       = {Daniel Beck},
  title        = {Reduce log levels for successful run completion and update center polling events by anonymous · Pull Request \#4345 · jenkinsci/jenkins},
  year         = {2019},
  date         = {2019-11-05},
  url          = {https://git.io/JJWzA},
  note         = {Following this URL may reveal author identity},
  organization = {Jenkins},
  urldate      = {2020-07-21},
}

@InProceedings{Leavens1990,
  author    = {Leavens, Gary T. and Weihl, William E.},
  title     = {Reasoning about Object-Oriented Programs That Use Subtypes},
  booktitle = ecoop,
  year      = {1990},
  series    = {OOPSLA/ECOOP ’90},
  publisher = {ACM},
  isbn      = {0897914112},
  pages     = {212--223},
  doi       = {10.1145/97945.97970},
  numpages  = {12},
}

@Article{Liskov1994,
  author     = {Liskov, Barbara H. and Wing, Jeannette M.},
  title      = {A Behavioral Notion of Subtyping},
  journal    = toplas,
  year       = {1994},
  volume     = {16},
  number     = {6},
  pages      = {1811--1841},
  issn       = {0164-0925},
  doi        = {10.1145/197320.197383},
  address    = {New York, NY, USA},
  issue_date = {Nov. 1994},
  numpages   = {31},
  publisher  = {ACM},
}

@InProceedings{Gharbi2019,
  author       = {Gharbi, Sirine and Mkaouer, Mohamed Wiem and Jenhani, Ilyes and Messaoud, Montassar Ben},
  title        = {On the Classification of Software Change Messages Using Multi-label Active Learning},
  booktitle    = {Symposium on Applied Computing},
  year         = {2019},
  series       = {SAC '19},
  organization = {ACM/SIGAPP},
  publisher    = {ACM},
  location     = {Limassol, Cyprus},
  isbn         = {978-1-4503-5933-7},
  pages        = {1760--1767},
  doi          = {10.1145/3297280.3297452},
  acmid        = {3297452},
  address      = {New York, NY, USA},
  numpages     = {8},
}

@InProceedings{Tian2017,
  author    = {Tian, Yuchi and Ray, Baishakhi},
  title     = {Automatically Diagnosing and Repairing Error Handling Bugs in C},
  booktitle = fse,
  year      = {2017},
  series    = {ESEC/FSE 2017},
  publisher = {ACM},
  location  = {Paderborn, Germany},
  isbn      = {978-1-4503-5105-8},
  pages     = {752--762},
  doi       = {10.1145/3106237.3106300},
  acmid     = {3106300},
  address   = {New York, NY, USA},
  numpages  = {11},
}

@InProceedings{Khatchadourian2020,
  author       = {Khatchadourian, Raffi and Tang, Yiming and Bagherzadeh, Mehdi and Ray, Baishakhi},
  title        = {An Empirical Study on the Use and Misuse of {Java} 8 Streams},
  booktitle    = fase,
  year         = {2020},
  editor       = {Wehrheim, Heike and Cabot, Jordi},
  series       = {FASE '20},
  organization = {ETAPS},
  publisher    = {Springer International Publishing},
  month        = apr,
  isbn         = {978-3-030-45234-6},
  pages        = {97--118},
  doi          = {10.1007/978-3-030-45234-6_5},
  address      = {Cham},
  data         = {http://doi.org/10.5281/zenodo.3677449},
  numpages     = {22},
  slides       = {https://www.slideshare.net/khatchad/an-empirical-study-on-the-use-and-misuse-of-java-8-streams-231309312},
}

@InProceedings{Mondal2017,
  author    = {Mondal, Manishankar and Roy, Chanchal K. and Schneider, Kevin A.},
  title     = {Identifying Code Clones Having High Possibilities of Containing Bugs},
  booktitle = {International Conference on Program Comprehension},
  year      = {2017},
  series    = {ICPC ’17},
  publisher = {IEEE Press},
  location  = {Buenos Aires, Argentina},
  isbn      = {9781538605356},
  pages     = {99--109},
  doi       = {10.1109/ICPC.2017.31},
  numpages  = {11},
}

@Dataset{Anonymous2020a,
  author    = {Anonymous},
  title     = {Automated Evolution of Feature Logging Statement Levels Using {Git} Histories and Degree of Interest},
  year      = {2020},
  month     = aug,
  doi       = {10.5281/zenodo.3698983},
  publisher = {Zenodo},
}

\end{document}